\documentclass[a4paper,prd,onecolumns,nofootinbib,preprintnumbers]{revtex4-2}

\usepackage{amsmath}
\usepackage{amssymb}
\usepackage[english]{babel}
\usepackage{braket}
\usepackage{cancel}
\usepackage{graphicx}
\usepackage{color}
\usepackage[utf8]{inputenc}
\usepackage{hyperref}
\usepackage{multirow}
\usepackage{slashed}
\usepackage[normalem]{ulem}
\usepackage{wasysym}
\usepackage{amssymb}
\usepackage{braket}
\usepackage{simplewick}
\usepackage{array}
\usepackage{graphicx}
\usepackage{float}
\usepackage{titlesec}
\usepackage{listings}
\usepackage{hyperref}
\usepackage[english]{babel}
\usepackage{xcolor}
\usepackage{xspace}
\usepackage{booktabs}

\hypersetup{colorlinks, linkcolor = [rgb]{0,0.0,0.75}, citecolor = [rgb]{0,0.0,0.75}, urlcolor = [rgb]{0,0.0,0.75}}

\newcommand{\mLam}{m_{\Lambda}}
\newcommand{\mLamB}{m_{\Lambda_b}}

\newcommand{\refeq}[1]{\mbox{Eq.~\eqref{eq:#1}}}
\newcommand{\refsec}[1]{\mbox{Sec.~\ref{sec:#1}}}
\newcommand{\reftab}[1]{\mbox{Table \ref{tab:#1}}}
\newcommand{\reffig}[1]{\mbox{Figure \ref{fig:#1}}}

\newcommand{\EOS}{\texttt{EOS}\xspace}

\newcolumntype{C}[1]{>{\centering\let\newline\\\arraybackslash\hspace{0pt}}m{#1}}

\begin{document}


\title{Dispersive bounds for local form factors in \texorpdfstring{$\Lambda_b \to \Lambda$}{Lb -> L} transitions}
\author{Thomas Blake}
\affiliation{Department of Physics, University of Warwick, Coventry, CV4\,7AL, UK}
\author{Stefan Meinel}
\affiliation{Department of Physics, University of Arizona, Tucson, AZ 85721, USA}
\author{Muslem Rahimi}
\affiliation{Center for Particle Physics Siegen (CPPS), Theoretische Physik 1, Universit{\"a}t Siegen, 57068 Siegen, Germany}
\author{Danny van Dyk}
\affiliation{Physik Department T31, Technische Universit\"at M\"unchen, 85748 Garching, Germany}

\begin{abstract}
We investigate the ten independent local form factors relevant to the $b$-baryon decay $\Lambda_b \to \Lambda \ell^+\ell^-$, combining information of lattice QCD and dispersive bounds. 
We propose a novel parametrization
of the form factors in terms of orthonormal polynomials that diagonalizes
the form factor contributions to the dispersive bounds.
This is a generalization of the
unitarity bounds developed for meson-to-meson form factors.
In contrast to {\it ad hoc} parametrizations of these form factors, 
our parametrization provides a degree of control of the form-factor uncertainties at large hadronic recoil.
This is of phenomenological interest for theoretical predictions of, {\it e.g.}, $\Lambda_b\to \Lambda \gamma$ and $\Lambda_b\to\Lambda \ell^+\ell^-$ decay processes. 

\end{abstract}

\preprint{EOS-2022-01, P3H-22-046, SI-HEP-2022-09, TUM-HEP 1399/22}

\maketitle

\section{Introduction}
\label{sec::Introduction}

For the last decade, decays involving $b\to s\mu^+\mu^-$ transitions have been a focus of flavour physics community due to the substantial
number of so-called ``$b$ anomalies.'' These anomalies are a pattern of deviations between theoretical expectations, within the Standard Model of
particle physics (SM), and experimental measurements, chiefly by the LHCb experiment~\cite{%
LHCb:2017avl,LHCb:2020lmf,LHCb:2021vsc,LHCb:2021zwz,LHCb:2021xxq,LHCb:2021trn%
}. Compatible experimental results, for many of these  measurements, have since been obtained
by the ATLAS~\cite{ATLAS:2018cur,ATLAS:2018gqc}, CMS~\cite{CMS:2019bbr,CMS:2017rzx,CMS:2020oqb}, and Belle~\cite{Belle:2016fev} experiments.

There is substantial interest in corroborating the $b$ anomalies through decay channels
that feature complementary sources of theoretical systematic uncertainties \emph{and} complementary sensitivity to  effects beyond the SM. The decay $\Lambda_b\to\Lambda(\to p \pi^-)\mu^+\mu^-$
is a prime candidate for this task~\cite{Boer:2014kda}.
In contrast to $B \to K^*(\to K\pi) \mu^+\mu^-$ decays, the local form factors for $\Lambda_b \to \Lambda\mu^+\mu^-$ decays correspond to transition matrix elements between stable single-hadron states in QCD. This allows precise lattice QCD calculations using standard methods, and results for the $\Lambda_b\to\Lambda$ form factors have been available for some time \cite{Detmold:2016pkz}.
Measurements of $\Lambda_b\to\Lambda(\to p \pi^-)\mu^+\mu^-$ observables \cite{LHCb:2015tgy,LHCb:2019wwi} have been included in global fits of the
$b\to s\mu^+\mu^-$ couplings~\cite{Blake:2019guk,Altmannshofer:2021qrr,Hurth:2020rzx,Bhom:2020lmk}, and dedicated analyses for effects beyond the SM, even accounting for production polarization
of the $\Lambda_b$, have been performed in recent years~\cite{Meinel:2016grj,Blake:2019guk}.
Lepton-flavor universality violation in baryonic $b \to s \ell^+ \ell^-$ decay modes has also been studied theoretically; in Ref.~\cite{Bordone:2021usz} the angular distribution of $\Lambda_b \to \Lambda \ell^+ \ell^-$ has been computed for the full base of New Physics operators (partial results are available in Refs.~\cite{Sahoo:2016nvx,Das:2019omf}). Measurements by LHCb are also available for the branching fraction of the $\Lambda_b \to \Lambda\gamma$ decay~\cite{LHCb:2019wwi}.

In this work, we investigate one of the two main sources of theoretical uncertainties that arise in the predictions
of $\Lambda_b\to \Lambda\ell^+\ell^-$ and $\Lambda_b \to \Lambda\gamma$ transitions; the hadronic form factors of local $\bar{s}\Gamma b$ currents of mass dimension three.
The complete set of scalar-valued hadronic form factors describing these currents is comprised of ten independent
functions of the dilepton invariant mass squared, $q^2$.
A convenient Lorentz decomposition of the hadronic matrix elements is achieved in terms of helicity amplitudes~\cite{Feldmann:2011xf}.
Here, we set out to improve the description of the form factors as functions of $q^2$ across the whole kinematic
phase space available to the  $\Lambda_b\to\Lambda\ell^+\ell^-$ decay. To that end, we derive dispersive bounds for the form factors in the 
six $\bar{s} \Gamma b$ currents: the (pseudo)scalar, the (axial)vector, and the two tensor currents.
We demonstrate that previous analyses of dispersive bounds for baryon-to-baryon form factors~\cite{%
Boyd:1995tg,Hill:2010yb,Bhattacharya:2011ah,Cohen:2019zev%
} overestimate the saturation of the bounds (see also the discussion in Ref.~\cite{Gambino:2020jvv}).
Our formulation of the bounds uses polynomials that are orthonormal on an arc of the unit circle in the variable $z$ (see Sec.~\ref{sec:th:parametrization} for the definition); such polynomials were previously used to parameterize the nonlocal matrix elements contributing to $B_{(s)}\to \{K^{(*)},\phi\}\ell^+\ell^-$ \cite{Gubernari:2020eft}.
As a consequence, benefits inherent to meson-to-meson form-factor parametrizations with dispersive bounds
now also apply to our approach. 
We illustrate the usefulness of our formulation of the dispersive bounds for the
form-factor parameters for $\Lambda_b\to\Lambda$, but note that it applies similarly to other ground-state baryon
to ground-state baryon form factors ({\it e.g.} $\Lambda_b \to \Lambda_c$ transitions). As inputs, we use lattice-QCD determinations of the form factors, which have already been extrapolated to the continuum limit and to physical quark masses, at up
to three different points in $q^2$. Our analysis also paves the way for
the application of the bounds directly, through a modified $z$-expansion, within future lattice QCD studies. This is likely to increase the precision of future form-factor predictions, especially at large hadronic recoil where $q^2\simeq 0$. 

In \refsec{th}, we briefly recap the theory of the local form factors for baryon-to-baryon transitions and their dispersive bounds.
We then propose a new parametrization for the full set of form factors in $\Lambda_b \to \Lambda$ transitions, which diagonalizes the dispersive bound. 
In Sec.~\ref{sec::Numerical-Analysis}, we illustrate the power of our parametrization based on lattice QCD constraints for the $\Lambda_b\to \Lambda$ form factors. 
We highlight how the form-factor uncertainties in the low momentum transfer region are affected by our parametrization
and the different types of bounds we apply. We conclude in \refsec{conc}.

\section{Derivation of the dispersive Bounds}
\label{sec:th}

We begin with a review of the Lorentz decomposition of the hadronic matrix elements in \refsec{th:lorentz}.
We then introduce the two-point correlation functions responsible for the dispersive bound and their theoretical
predictions within an operator product expansion in \refsec{th:db}.
The hadronic representation of the correlation functions is discussed in \refsec{th:had-repr}. Our proposed parametrization is introduced in \refsec{th:parametrization}.

\subsection{Lorentz decomposition in terms of helicity form factors}
\label{sec:th:lorentz}

A convenient definition of the form factors is achieved when each helicity amplitude corresponds to a single
form factor:
\begin{equation}
    \bra{\Lambda(k)} \bar{s} \Gamma^\mu b \ket{\Lambda_b(p)} \, \varepsilon^*_\mu(\lambda) \propto f^{\Gamma}_\lambda(q^2)\,,
\end{equation}
where $q^2 = (p - k)^2$, and $\varepsilon$ is the polarization vector of a fictitious vector mediator with
polarization $\lambda$.
For $1/2^+\to 1/2^+$ transitions, this definition is achieved by the Lorentz decomposition~\cite{Feldmann:2011xf}:
\begin{align}
    \label{eq:th:lorentz-decomposition:V}
    \bra{ \Lambda(k,s_\Lambda) } \overline{s} \,\gamma^\mu\, b \ket{ \Lambda_b(p,s_{\Lambda_{b}}) }
        & = \overline{u}_\Lambda(k,s_{\Lambda}) \bigg[ f_t^V(q^2)\: (m_{\Lambda_b}-m_\Lambda)\frac{q^\mu}{q^2} \\
    \nonumber
        &   \phantom{\overline{u}_\Lambda \bigg[}+ f_0^V(q^2) \frac{m_{\Lambda_b}+m_\Lambda}{s_+}
            \left( p^\mu + k^{ \mu} - (m_{\Lambda_b}^2-m_\Lambda^2)\frac{q^\mu}{q^2}  \right) \\
    \nonumber
        &   \phantom{\overline{u}_\Lambda \bigg[}+ f_\perp^V(q^2)
            \left(\gamma^\mu - \frac{2m_\Lambda}{s_+} p^\mu - \frac{2 m_{\Lambda_b}}{s_+} k^{ \mu} \right) \bigg] u_{\Lambda_b}(p,s_{\Lambda_{b}}) \, , \\
    \label{eq:th:lorentz-decomposition:A}
    \bra{ \Lambda(k,s_{\Lambda}) } \overline{s} \,\gamma^\mu\gamma_5\, b \ket{ \Lambda_b(p,s_{\Lambda_{b}}) }
        & = -\overline{u}_\Lambda(k,s_{\Lambda}) \:\gamma_5 \bigg[ f_t^A(q^2)\: (m_{\Lambda_b}+m_\Lambda)\frac{q^\mu}{q^2} \\
    \nonumber
        &   \phantom{\overline{u}_\Lambda \bigg[}+ f_0^A(q^2)\frac{m_{\Lambda_b}-m_\Lambda}{s_-}
            \left( p^\mu + k^{ \mu} - (m_{\Lambda_b}^2-m_\Lambda^2)\frac{q^\mu}{q^2}  \right) \\
    \nonumber 
        & \phantom{\overline{u}_\Lambda \bigg[}+ f_\perp^A(q^2) \left(\gamma^\mu + \frac{2m_\Lambda}{s_-} p^\mu - \frac{2 m_{\Lambda_b}}{s_-} k^{ \mu} \right) \bigg]  u_{\Lambda_b}(p_{\Lambda_{b}},s_{\Lambda_{b}}), \\
    \bra{ \Lambda(k,s_{\Lambda}) } \overline{s} \,i\sigma^{\mu\nu} q_\nu \, b \ket{ \Lambda_b(p,s_{\Lambda_{b}}) } 
        &= - \overline{u}_\Lambda(k,s_{\Lambda}) \bigg[  f_0^T(q^2) \frac{q^2}{s_+} \left( p^\mu + k^{\mu} - (m_{\Lambda_b}^2-m_{\Lambda}^2)\frac{q^\mu}{q^2} \right) \\
    \nonumber 
        & \phantom{\overline{u}_\Lambda \bigg[} + f_\perp^T(q^2)\, (m_{\Lambda_b}+m_\Lambda) \left( \gamma^\mu -  \frac{2  m_\Lambda}{s_+} \, p^\mu - \frac{2m_{\Lambda_b}}{s_+} \, k^{ \mu}   \right) \bigg] u_{\Lambda_b}(p,s_{\Lambda_{b}}) \, , \\
    \label{eq:th:lorentz-decomposition:T}
    \bra{ \Lambda(k,s_\Lambda) } \overline{s} \, i\sigma^{\mu\nu}q_\nu \gamma_5  \, b \ket{ \Lambda_b(p,s_{\Lambda_{b}}) }
        & = -\overline{u}_{\Lambda}(k,s_{\Lambda}) \, \gamma_5 \bigg[   
        f_0^{T5}(q^2) \, \frac{q^2}{s_-}
            \left( p^\mu + k^{\mu} -  (m_{\Lambda_b}^2-m_{\Lambda}^2) \frac{q^\mu}{q^2} \right) \\
    \nonumber
        &   \phantom{\overline{u}_\Lambda \bigg[}  + f_\perp^{T5}(q^2)\,  (m_{\Lambda_b}-m_\Lambda)
            \left( \gamma^\mu +  \frac{2 m_\Lambda}{s_-} \, p^\mu - \frac{2 m_{\Lambda_b}}{s_-} \, k^{ \mu}  \right) \bigg]  u_{\Lambda_b}(p,s_{\Lambda_{b}})\,,
\end{align}
where we abbreviate $\sigma^{\mu \nu} = \frac{i}{2} [\gamma^\mu, \gamma^\nu]$
and $s_{\pm} = (\mLamB \pm \mLam)^2 - q^2$. The labelling of the ten form factors follows the conventions of
Ref.~\cite{Boer:2014kda}. Each form factor, $f_\lambda^\Gamma$, arises in the current $\bar{s} \Gamma b$ in a helicity amplitude
with polarization $\lambda = t, 0, \perp$. We refer to Ref.~\cite{Boer:2014kda} for details and the relations between
the form factors and the helicity amplitudes.
Note that the matrix elements for the scalar and pseudo-scalar current can be related to the vector and axial-vector current of the timelike-polarized form factors $f_t^V$ and $f_t^A$ via the equations of motion:
\begin{align}
    \bra{ \Lambda(k,s_\Lambda) } \overline{s}  \, b \ket{ \Lambda_b(p,s_{\Lambda_{b}}) } 
    \nonumber
        &= \frac{q^\mu}{m_b -m_s} \bra{ \Lambda(k,s_\Lambda) } \overline{s} \,\gamma_\mu  \, b \ket{ \Lambda_b(p,s_{\Lambda_{b}}) } \\
        &= f_t^V(q^2) \frac{\mLamB - \mLam}{m_b -m_s}  \overline{u}_\Lambda(k,s_\Lambda)  \, u_{\Lambda_b}(p,s_{\Lambda_{b}}) \, ,\\
    \bra{ \Lambda(k,s_\Lambda) } \overline{s}  \, \gamma_5 \, b \ket{ \Lambda_b(p,s_{\Lambda_{b}}) } 
    \nonumber 
        &= -\frac{q^\mu}{m_b + m_s} \bra{ \Lambda(k,s_\Lambda) } \overline{s} \,\gamma_\mu \gamma_5 \, b \ket{ \Lambda_b(p,s_{\Lambda_{b}}) } \\
        &= f_t^A(q^2) \frac{\mLamB + \mLam}{m_b + m_s} \overline{u}_\Lambda(k,s_\Lambda)  \, \gamma_5 \, u_{\Lambda_b}(p,s_{\Lambda_{b}}) \, .
\end{align}

Although the ten functions, $f_\lambda^\Gamma(q^2)$, are {\it a priori} independent, some relations exist at
specific points in $q^2$. These so-called endpoint relations arise due to two different mechanisms.
First, the hadronic matrix elements on the left-hand sides of \refeq{th:lorentz-decomposition:V} to \refeq{th:lorentz-decomposition:T}
must be free of kinematic singularities. Two such 
singularities can arise, as spurious poles at $q^2 = 0$ and $q^2 = q^2_\text{max} \equiv (\mLamB - \mLam)^2$.
They are removed by the following identities:
\begin{align}
    \label{eq:ep1}
    f_t^V(0)
        & = f_0^V(0)  \, , &
    f_t^A(0)
        & = f_0^A(0)\,,\\
    \label{eq:ep2}
    f_\perp^A(q^2_{\text{max}})
        & = f_0^A(q^2_{\text{max}}) \, ,&
    f_\perp^{T5}(q^2_{\text{max}})
        & = f_0^{T5}(q^2_{\text{max}})\,.
\end{align}
In addition to the above, an algebraic relation between $\sigma^{\mu\nu}$ and $\sigma^{\mu\nu}\gamma_5$
ensures that
\begin{align}
    \label{eq:ep3}
    f_\perp^{T5}(0)
        & = f_\perp^T(0)\,.
\end{align}
See also Ref.~\cite{Hiller:2021zth} for additional discussion of endpoint relations for baryon transition form factors.

\subsection{Two-point correlation functions and OPE representation}
\label{sec:th:db}

\begin{figure}
    \centering
    \includegraphics[width=.7\textwidth]{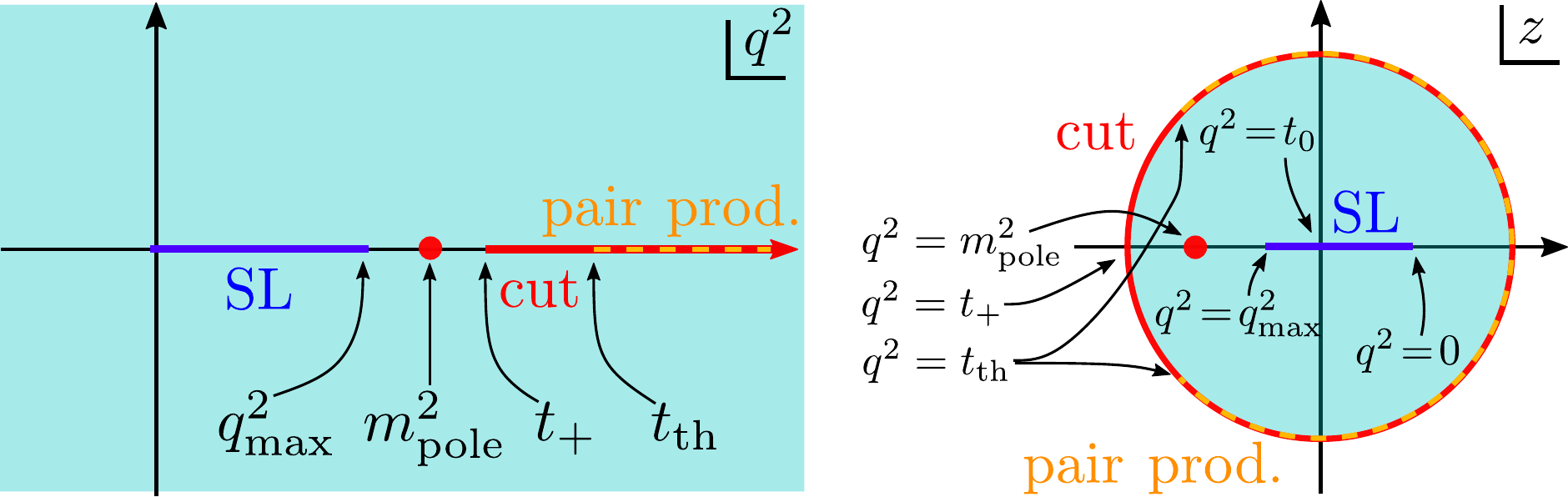}
    \caption{%
    Sketch of the analytic structure of the baryon-to-baryon form factors in the variable $q^2$ (left)
    and the variable $z$ (right). The $q^2$ range of semileptonic decays is marked ``SL''.
    The baryon/antibaryon pair production is marked ``pair prod.''. The form factors develop a branch cut
    below the baryon/antibaryon pair production threshold due to rescattering of virtual baryon/antibaryon
    pairs into, {\it e.g.}, $\bar{B}K^{(*)}$ pairs.
    }
    \label{fig:th:sketch}
\end{figure}

Dispersive bounds for local form factors have a successful history. They were first
used for the kaon form factor \mbox{\cite{Okubo:1971jf,Okubo:1971my,Okubo:1971wup}} and have also successfully been applied to
exclusive $B\to \pi$~\cite{Becher:2005bg,Bourrely:2008za} and $B\to D^{(\ast)}$~\cite{Boyd:1994tt,Boyd:1995tg,Boyd:1997qw} form factors\footnote{See also applications \cite{Lellouch:1995yv,DiCarlo:2021dzg,Martinelli:2021frl,Martinelli:2021onb,Martinelli:2022vvh} of the dispersive matrix method \cite{Caprini:2019osi}.}.
In the latter case, the heavy-quark expansion renders the bounds phenomenologically
more useful due to relations between all form factors of transitions between doublets
under heavy-quark spin symmetry~\cite{Caprini:1997mu}, see  Refs.~\cite{Bigi:2017jbd,Bordone:2019guc,Bordone:2019vic}
for recent phenomenological updates and analyses up to order $1/m^2$ in the heavy-quark expansion, respectively.
The application of the bound to form factors arising in baryon-to-baryon transitions is more complicated~\cite{Hill:2010yb,Gambino:2020jvv},
chiefly due to the fact that for any form factor, $F$, its first branch point, $t_+^F$, does not coincide with the
threshold for baryon/antibaryon pair production, $t_{\rm th}^F$. Instead, the branch points lay to the left
of the pair production points, at the pair production threshold for the corresponding ground-state meson/antimeson pair.
We show a sketch of this structure in the left-hand side of \reffig{th:sketch}.

The dispersive bounds connect a theoretical computation of
a suitably-chosen two-point function with weighted integrals of the squared hadronic form factors.
For concreteness and brevity we derive the dispersive bound for the vector current $J_V^\mu$
and its hadronic form factors. The generalization to the currents
\begin{align}
    J^\mu_{V}
        & = \bar{s} \gamma^\mu b \, ,&
    J^\mu_{A}
        & = \bar{s} \gamma^\mu \gamma_5 b \, ,\\
    J^{\mu}_{T}
        & = \bar{s} \sigma^{\mu \nu} q_\nu b \, ,&
    J^{\mu}_{T5}
        & = \bar{s} \sigma^{\mu \nu}  q_\nu \gamma_5 \,b
\end{align}
is straightforward following the same prescription as $J_V^{\mu}$. 
As we will see below, the results for scalar and pseudo-scalar currents can be obtained
from the vector and axial currents, respectively.

We define $\Pi^{\mu\nu}_V$ to be the vacuum matrix elements of the two-point function with two insertions of $J_V$:
\begin{align}
    \label{eq:th:db:Pi_mu_nu}
     \Pi^{\mu \nu}_{V}(Q)
        &= i \int \text{d}^4 x \, \, e^{i Q\cdot x} \bra{0} \mathcal{T} \{ J_V^{\mu}(x), J_V^{\nu \dagger}(0) \} \ket{0}\,,
\end{align}
where $Q^\mu$ is the four-momentum flowing through the two-point function.
This tensor-valued function can be expressed in terms of two scalar-valued functions:
\begin{align}
    \label{eq:th:db:Pi}
    \Pi_{V}^{\mu\nu}(Q) &= P^{\mu \nu}_{J=0}(Q) \Pi_{V}^{J=0}(Q^2) + 3\,P^{\mu\nu}_{J=1}(Q) \Pi_V^{J=1}(Q^2)\,,
\end{align}
using the two projectors
\begin{align}
    P^{\mu \nu}_{J=0}(p)
        & = \frac{p^\mu p^\nu}{p^2} \, ,&
    P^{\mu\nu}_{J=1}(p)
        & = \frac{1}{3} \left(\frac{p^\mu p^\nu}{p^2} - g^{\mu \nu} \right) \, .
\end{align}
Note that the two tensor currents do not feature a $J=0$ component, {\it i.e.}, the coefficients of the 
projectors $P_{J=0}$ vanish for these currents.

The functions $\Pi_V^{J=0}(Q^2)$ and $\Pi_V^{J=1}(Q^2)$ feature singularities along the real $Q^2$ axis,
which will be discussed below. These singularities are captured by the discontinuities of
$\Pi_V^{J=0}$ and $\Pi_V^{J=1}$. It is now convenient to define a new function, $\chi^J_V$,
which is completely described in terms of the discontinuities of the functions $\Pi^{J=1}_V$:
\begin{equation}
    \label{eq:th:db:def-chi}
    \chi^{J=1}_V(Q^2)
        = \frac{1}{n!} \left(\frac{d}{dQ^2}\right)^n \Pi_V^{J=1}(Q^2)
        = \frac{1}{2 \pi i} \int_{0}^\infty \text{d}t \, \frac{\text{Disc} \, \Pi^{J=1}_V(t)}{(t - Q^2)^{n + 1}}\,.
\end{equation}
Here, the number of derivatives $n$ (also known as the number of ``subtractions'') is chosen to be the smallest number that yields a convergent integral.
Note that in general the functions $\chi$ for the scalar and pseudo-scalar currents require a different value of $n$ than the
functions for the vector and axial currents, respectively, despite the fact that they can be extracted from the
vector and axial two-point correlators.

\begin{table}[t]
    \begin{center}
    \renewcommand{\arraystretch}{1.25}
    \begin{tabular}{C{1cm} C{1cm} C{3cm} C{4cm} C{1cm}}
        \toprule
        $\Gamma$ & $J$ & form factors               & $\chi_{\Gamma}^{J}|_\text{OPE}$ [$10^{-2}$]
                                                                            & $n$ \\
        \midrule
        $V$      & $0$ & $f_t^V$                    & $1.42$                & $1$ \\
        $V$      & $1$ & $f_0^V$, $f_\perp^V$       & $1.20 \, / \, m_b^2$  & $2$ \\
        $A$      & $0$ & $f_t^A$                    & $1.57$                & $1$ \\
        $A$      & $1$ & $f_0^A$, $f_\perp^A$       & $1.13 \, / \, m_b^2$  & $2$ \\
        $T$      & $1$ & $f_0^T$, $f_\perp^T$       & $0.803 \, / \, m_b^2$ & $3$ \\
        $T5$     & $1$ & $f_0^{T5}$, $f_\perp^{T5}$ & $0.748 \, / \, m_b^2$ & $3$ \\
        \bottomrule
    \end{tabular}
    \renewcommand{\arraystretch}{1.0}
    \end{center}
    \caption{%
        \label{tab:th:db:chiOPE-and-n}
        The values of $\chi_{\Gamma}^{J}(Q^2 = 0)|_\text{OPE}$ as taken from Ref.~\cite{Bharucha:2010im},
        which include terms at next-to-leading order in $\alpha_s$ and subleading power corrections.
        The number of derivatives for each current $\Gamma = V,A,S,P,T,T5$ is provided as $n$.
        Note that the results for $\chi$ in the rows for $\Gamma = T,T5$ differ from
        those given in Ref.~\cite{Bharucha:2010im} by a factor of $\tfrac{1}{4}$, which is
        due to differences in convention for the tensor current.
        The value of the $b$-quark mass is taken as $m_b = 4.2$ GeV.
    }
\end{table}

The dispersive bound is constructed by equating two different representations of $\chi_V$ with each other,
based on the assumption of global \emph{quark hadron duality}:
\begin{equation}
    \label{eq:th:db:QHD}
    \chi_{V}^{J}\bigg|_\text{OPE} =\chi_{V}^{J}\bigg|_\text{hadr}\,.
\end{equation}
The left-hand side representation is obtained from an operator product expansion (OPE) of the time-ordered product that gives rise to $\Pi_V^{\mu\nu}(Q)$. For $\bar{s}\Gamma b$ currents,
the most recent analysis of these OPE results, including subleading contributions,
has been presented in Ref.~\cite{Bharucha:2010im} for all the dimension-three currents considered
in this work. We summarize results of the analysis for $Q^2 = 0$ in \reftab{th:db:chiOPE-and-n}, where we
also list the values for $n$ on a per-current basis.\\
The right-hand side representation is obtained from the hadronic matrix elements of on-shell intermediate states.
We will discuss this representation and its individual terms in the next section.

\subsection{Hadronic representation of the bound}
\label{sec:th:had-repr}

We continue to discuss the bounds for the case of the vector current, and concretely,
the scalar-valued two-point function $\Pi_V^{J=1}$,
\begin{equation}
    \Pi_V^{J=1} = \left[P_{J=1}\right]_{\mu\nu} \Pi^{\mu\nu}_V\,.
\end{equation}
Its discontinuity due to a hadronic intermediate state, $H_{\bar{s}b}$, with flavour quantum numbers
$B = -S = 1$ can be obtained using
\begin{align}
    \label{eq:th:had-repr:def-disc:Disc-oneparticle}
    \text{Disc} \,\Pi_{\Gamma}^{J}
        & = i \sum_\textrm{spin} \int \text{d} \rho \, (2\pi)^4 \delta^{(4)}\left(q- \sum_i^n p_i\right) P_{J}^{\mu \nu}(q) \bra{0} J^{\mu}_{\Gamma} \ket{H_{b\bar{s}}(p_1, \dots, p_n)}  \bra{H_{b\bar{s}}(p_1, \dots, p_n)} J^{\nu \dagger}_{\Gamma} \ket{0}\,,
\end{align}
where the $\text{d}\rho$ is the phase-space element of the $n$-particle intermediate state. Below we consider the cases of one- and two-particle intermediate states, with
\begin{align}
 \label{eq::phasespace-integral}
 \int \text{d}\rho  &= 
  \begin{cases}
    \displaystyle\int \dfrac{\text{d}^3 p}{(2\pi)^3 2 E_{\vec{p}}}
        &   \text{for one-particle states}, \vspace{0.25cm}\\
    \displaystyle\int \dfrac{\text{d}^3 p_1}{(2\pi)^3 2 E_{\vec{p}_1}} \displaystyle\int \dfrac{\text{d}^3 p_2}{(2\pi)^3 2 E_{\vec{p}_2}}
        &   \text{for two-particle states}.
  \end{cases}
\end{align}

\subsubsection{One-particle contributions}
\label{sec:one-particle}
Here, we discuss contributions due to a single asymptotic on-shell state $H_{b\bar{s}}$ with flavour
quantum numbers $B = -S = 1$, which excludes states that strongly decay such as radially excited states.
We continue to use the case $\Gamma=V$ as an example, with $J=1$. In that case, the discontinuity
receives a single contribution:
\begin{align}
    \text{Disc} \, \Pi_{V}^{J=1}(q^2)\bigg|_{\text{1pt}}
        & = i \int \text{d}\rho \,  (2\pi)^4 \delta^{(4)}(q-p) \sum_\lambda \left[P_{J=1} \right]_{\mu \nu} \bra{0} J^{\mu}_{V} \ket{\bar{B}^{*}_{s}(p, \lambda)}  \bra{\bar{B}^{*}_{s}(p, \lambda)} J^{\nu \dagger}_{V} \ket{0}\\
        & = i \int \text{d}\rho \,  (2\pi)^4 \delta^{(4)}(q-p) m_{B_s^*}^2 f_{B_s^*}^2\\
        & = 2\pi \delta(q^2 - m_{B_s^*}^2) \theta(q^0) m_{B_s^*}^2 f_{B_s^*}^2\,,
    \label{eq:Disc-V-1pt}
\end{align}
where $\lambda$ is the polarization of the $\bar{B}_s^*$ meson and $m_{B_s^*}$ its mass. States other than the $\bar{B}_s^*$ do not contribute,
since either their matrix elements with the $\Gamma = V$ current vanish, their projection onto the $J=1$ state vanishes,
or they decay strongly.
The generalization to $\Gamma = A$ and $J=0$ is straightforward:
\begin{align}
    \text{Disc} \, \Pi_{V}^{J=0}(q^2)\bigg|_{\text{1pt}}
        & = 2\pi \delta(q^2 - m_{B^*_{s,0}}^2) \theta(q^0) m_{B^*_{s,0}}^2 f_{B^*_{s,0}}^2 \,, \\
    \text{Disc} \, \Pi_{A}^{J=0}(q^2)\bigg|_{\text{1pt}}
        & = 2\pi \delta(q^2 - m_{B_s}^2) \theta(q^0) m_{B_s}^2 f_{B_s}^2\,, \\
    \text{Disc} \, \Pi_{A}^{J=1}(q^2)\bigg|_{\text{1pt}}
        & =  2\pi \delta(q^2 - m_{B_{s,1}}^2) \theta(q^0) m_{B_{s,1}}^2 f_{B_{s,1}}^2 \,.
\end{align}
Here $B_s$ is the ground-state pseudoscalar meson with a very well-known decay constant
$f_{B_s} = 230.7 \pm 1.3\,\, \text{MeV}$~\cite{Bazavov:2017lyh},
$B_{s,1}$ is the axial vector meson, and $B_{s,0}^*$ is the scalar meson.
In brief, the (pseudo)scalar current receives a contribution from a (pseudo)scalar on-shell state, and the axialvector
current receives a contribution from an axialvector on-shell state.
Although sub-$BK$-threshold $B_{s,1}$ or $B_{s,0}^*$ states have not yet been seen in the experiment,
there are indications in lattice QCD analyses that such sub-threshold states exist \cite{Lang:2015hza}. However, the values
of their respective decay constants are presently not very well known; estimates have been obtained, via QCD sum rule at next-to-leading order, in Refs.~\cite{Gelhausen:2013wia,Pullin:2021ebn}.
Nevertheless, these states produce a pole both in the two-point functions $\Pi_{\Gamma}^J$ and in their associated form factors,
which is a necessary information for the formulation of the dispersive bounds and the form-factor parametrization.
From this point forward, we assume the presence of a single pole due to a $J^P=\lbrace 0^+, 1^-,0^-,1^+\rbrace$
state contributing to form factors with $(\Gamma,J) = \lbrace (V,0), (V,1), (A,0), (A,1)\rbrace$, respectively.

The cases for currents with $\Gamma = T$ and $\Gamma = T5$ benefit from further explanation. For these currents
one might assume that tensor, {\it i.e.}, $J^P=2^{\pm}$, states play a leading role. However, these states do not contribute
at all, since their matrix elements vanish:
\begin{equation}
    \bra{0} \bar{s} \sigma^{\mu\nu} (\gamma_5) b \ket{\bar{B}_s(J^P=2^\pm)} = 0\,.
\end{equation}
This can readily be understood, since the above matrix elements are antisymmetric in the
indices $\mu$ and $\nu$, while the polarization tensors of $J^P=2^\pm$ mesons are symmetric quantities.
Nevertheless, the currents $\Gamma = T$ and $\Gamma = T5$ do feature poles due to one-particle contributions,
which arise from states with $J^P=1^\pm$. We obtain: 
\begin{align}
    \text{Disc} \, \Pi_{T}^{J=1}(q^2)\bigg|_{\text{1pt}}
        & =  2\pi \delta(q^2 - m_{B_s^*}^2) \theta(q^0) m_{B_s^*}^4 (f_{B_s^*}^T)^2 \,, \\
    \text{Disc} \, \Pi_{T5}^{J=1}(q^2)\bigg|_{\text{1pt}}
        & =  2\pi \delta(q^2 - m_{B_{s,1}}^2) \theta(q^0) m_{B_{s,1}}^4 (f_{B_{s,1}}^T)^2 ,
\end{align}
where $f_{B_s^*}^{T}$ and $f_{B_{s,1}}^{T}$ are the decay constants of the respective state for a tensor
current:
\begin{align}
    \bra{0} J_{T}^\mu \ket{\bar{B}^*_{s}(p)} &= i m_{B_s^*}^2 f^T_{B_s^*} \epsilon^\mu \, &  \bra{0} J_{T5}^\mu \ket{\bar{B}_{s,1}(p)} &= - i m_{B_{s,1}}^2 f^T_{B_{s,1}} \epsilon^\mu \,.
\end{align}

Plugging the results for the discontinuities into \refeq{Disc-V-1pt} we obtain:
\begin{align}
    \chi_{V}^{J=1}(Q^2)\bigg|_\text{1pt}
        & = \frac{m_{B_s^*}^2 f_{B_s^*}^2}{(m_{B_s^*}^2 - Q^2)^{n+1}} \,, &
    \chi_{V}^{J=0}(Q^2)\bigg|_\text{1pt}
        & = \frac{m_{B^*_{s,0}}^2 f_{B^*_{s,0}}^2}{(m_{B^*_{s,0}}^2 - Q^2)^{n+1}} \,, \\
    \chi_{A}^{J=1}(Q^2)\bigg|_\text{1pt}
        & = \frac{m_{B_{s,1}}^2 f_{B_{s,1}}^2 }{(m_{B_{s,1}}^2 -Q^2)^{n+1}} \,, &
    \chi_{A}^{J=0}(Q^2)\bigg|_\text{1pt}
        & = \frac{m_{B_s}^2 f_{B_s}^2 }{(m_{B_s}^2 -Q^2)^{n+1}} \,, \\
    \chi_{T}^{J=1}(Q^2)\bigg|_\text{1pt}
        & = \frac{m_{B^*_s}^4 (f^T_{B_s^*})^2 }{(m_{B^*_s}^2 -Q^2)^{n+1}} \,, &
    \chi_{T5}^{J=1}(Q^2)\bigg|_\text{1pt}
        & = \frac{m_{B_{s,1}}^4 (f^T_{B_{s,1}})^2 }{(m_{B_{s,1}}^2 -Q^2)^{n+1}} \,.
\end{align}
The one-particle
contributions each amount to about $10\%$ of the respective OPE result.
\\

\subsubsection{Two-particle contributions}
\label{sec:two-particle}
Here, we focus on the contributions to $\chi$ due to an intermediate $\Lambda_b\bar{\Lambda}$ state.
By means of unitarity we can express the discontinuity of the two-particle correlator $\Pi^{J}_{\Gamma}(t)$ as a sum of intermediate $H_{b \bar{s}}$ states with flavour quantum numbers $B = -S= 1$:
\begin{align}
    \text{Disc} \, \Pi^{J}_{\Gamma}  &= i \sum_\textrm{spins} \int \text{d}\rho  \, (2\pi)^4 \delta^{(4)}(q - \left( p_1 + p_2 \right)) [P_{J}]_{\mu \nu} \bra{0} J^{\mu}_{\Gamma} \ket{\Lambda_b(p_1, s_{\Lambda_{b}}) \bar{\Lambda}(-p_2,s_\Lambda)}\nonumber \\
    & \times  \bra{\bar{\Lambda}(-p_2,s_\Lambda) \Lambda_b(p_1, s_{\Lambda_{b}})} J^{\nu \dagger}_{\Gamma} \ket{0} + \text{further positive terms} \, .
\end{align}
Note that further two-particle contributions for which dispersive bounds have been applied include
$\bar{B} K, \bar{B} K^*$, and $\bar{B}_s \phi$~\cite{Bharucha:2010im}.
The effect of each of those two-particle contributions would decrease the upper bound only by \mbox{1--4\%}~\cite{Bharucha:2010im},
{\it i.e.}, by a smaller amount than the one-particle contributions.

We can evaluate the phase-space integration in the rest frame of the two-particle system as
\begin{align}
    \int \text{d}\rho  \, (2\pi)^4 \delta^{(4)}(q - \left( p_1 + p_2 \right))  &= \frac{1}{8 \pi} \frac{\sqrt{\lambda(\mLamB^2, \mLam^2, q^2)}}{q^2} \theta(q^2 - s_{\Lambda_{b} \Lambda}),
\end{align}
with $s_{\Lambda_{b} \Lambda} = (\mLamB + \mLam)^2$. From this we obtain
\begin{align}
    \text{Disc} \, \Pi^{J}_{\Gamma} &=  \frac{i}{8\pi} \frac{\sqrt{ \lambda(\mLamB^2, \mLam^2, q^2) }}{q^2} \theta(q^2-s_{\Lambda_b \Lambda})   [P_{J}]_{\mu \nu} \bra{0} J^{\mu}_{\Gamma} \ket{\Lambda_b \bar{\Lambda}} \bra{\bar{\Lambda} \Lambda_b} J^{\nu \dagger}_{\Gamma} \ket{0}
\end{align}
where in the last line we dropped all further positive terms. 
In the following we summarize the contraction between helicity operators and matrix elements that can be expressed via local form factors:

\begin{align}
     [P_{J}]_{\mu \nu} \bra{0} J^{\mu}_{V} \ket{\bar{\Lambda} \Lambda_b}  \bra{\bar{\Lambda} \Lambda_b} J^{\nu \dagger}_{V} \ket{0}  &= 
  \begin{cases}
    \dfrac{2(\mLamB-\mLam)^2}{q^2} s_+(q^2) |f_{t}^V|^2 & \text{for } J = 0, \vspace{0.25cm}\\ 
    \dfrac{2 s_{-}(q^2)}{3 q^2} \left( (\mLamB+ \mLam)^2 |f_{0}^V|^2 + 2 q^2 \, |f_{\perp}^V|^2 \right) & \text{for } J = 1,
  \end{cases}  \label{eq::vector-FF} \\
    [P_{J}]_{\mu \nu}\bra{0} J^{\mu}_{A} \ket{\bar{\Lambda} \Lambda_b}  \bra{\bar{\Lambda} \Lambda_b} J^{\nu \dagger}_{A} \ket{0}  &= 
  \begin{cases}
    \dfrac{2 s_{-}(q^2)}{q^2} (\mLamB+\mLam)^2 |f_{t}^A|^2 & \text{for } J = 0, \vspace{0.25cm}\\
    \dfrac{2 s_+(q^2)}{3 q^2} \left( (\mLamB-\mLam)^2 |f_{0}^A|^2 + 2 q^2 \, |f_{\perp}^A|^2 \right) & \text{for } J = 1,
  \end{cases}\\
      [P_{J}]_{\mu \nu} \bra{0} J^{\mu}_{T} \ket{\bar{\Lambda} \Lambda_b}  \bra{\bar{\Lambda} \Lambda_b} J^{\nu \dagger}_{T} \ket{0}  &= 
  \begin{cases}
   0 & \text{for } J = 0, \vspace{0.25cm}\\
   \dfrac{2 s_{-}(q^2)}{3}  \left( 2 (\mLamB+\mLam)^2 |f_{\perp}^T|^2 + q^2 \, |f_{0}^T|^2  \right) & \text{for } J = 1,
  \end{cases} \label{eq::axialvector-FF} \\
   [P_{J}]_{\mu \nu} \bra{0} J^{\mu}_{T5} \ket{\bar{\Lambda} \Lambda_b}  \bra{\bar{\Lambda} \Lambda_b} J^{\nu \dagger}_{T5} \ket{0}  &= 
  \begin{cases}
   0 & \text{for } J = 0, \vspace{0.25cm}\\
   \dfrac{2 s_{+}(q^2)}{3} \left( 2 (\mLamB-\mLam)^2 |f_{\perp}^{T5}|^2 + q^2 \, |f_{0}^{T5}|^2  \right) & \text{for } J = 1,
  \end{cases}
  \label{eq::tensor-FF}
\end{align}
where the sum over the baryon spins is implied.

\subsection{Parametrization}
\label{sec:th:parametrization}
We relate the OPE representation to the hadronic representation of the functions $\chi_\Gamma^J$ through \refeq{th:db:QHD}.
Using $\Gamma = V$ and $J=1$ again as an example, the dispersive bound takes the form
\begin{align}
    \label{eq:th:parametrization:bound-V1}
    \chi^{J=1}_{V}(Q^2)\bigg|_\text{OPE}
        & \geq \chi^{J=1}_{V}(Q^2)\bigg|_\text{1pt}
          + \int_{s_{\Lambda_b \Lambda} }^{\infty} \text{d}t \, \frac{1}{24 \pi^2} \frac{\sqrt{\lambda(\mLamB^2,\mLam^2,t) }}{t^2 (t-Q^2)^{n+1}} s_{-}(t) \\
          \nonumber 
          & \hspace{3cm} \times \left( (\mLamB +\mLam)^2 |f_{0}^V(t)|^2 +2 t |f_{\perp}^V(t)|^2  \right) \,,
\end{align}
where the last term is the two-particle contribution due to the ground-state baryons. Our intent is now to parametrize the $\Lambda_b\to\Lambda$ form factors (here: $f_0^V, f_\perp^V$)
in such a way that their parameters enter the two-particle contributions to $\chi_\Gamma$ in a simple form.
Concretely, we envisage a contribution that enters as the 2-norm of the vector of parameters.\\

In general, the bounds are best represented by transforming the variable $t$ to the new variable $z$, defined as
\begin{align}
    z(t; t_0, t_+) & = \frac{\sqrt{t_+ - t} -\sqrt{t_+ - t_0}}{\sqrt{t_+ - t} +\sqrt{t_+ - t_0}}\,.
\end{align}
In the above, $t_0$ corresponds to the zero of $z(t)$ and is a free parameter that can be chosen, and $t_+$ corresponds to lowest branch point of the form factors.
The mapping from $t=q^2$ to $z$ is illustrated in \reffig{th:sketch}. The integral comprising the two-particle contribution starts at the pair-production threshold $t_\text{th}$. \\

When discussing the dispersive bounds for e.g.~$B\to D$ or $B\to\pi$ form factors, one has  $t_\text{th}=t_+$. The integral of the discontinuity along the real $t$ axis in the mesonic analogue of~\refeq{th:parametrization:bound-V1} then becomes a contour integral along the unit circle $|z|=1$. For an arbitrary function $g$, \begin{equation}
    \int_{t_\text{th} = t_+}^\infty \text{d}t \,\text{Disc} \, g(t)
        = \frac{1}{2} \oint_{|z| = 1} \frac{\text{d}z}{z} \left|\frac{\text{d}t(z)}{\text{d}z}\right| \,\text{Disc} \,g(t(z))
        = \frac{i}{2} \int_{-\pi}^{+\pi} \text{d}\alpha \,  \left| \frac{\text{d}t(z)}{\text{d}z}\right| \, \text{Disc} \,g(t(e^{i\alpha}))\,.
\end{equation}
The contribution to the integrand from a form factor $F$ is then written as $|\phi_F|^2 |F|^2$, where the \emph{outer function} $\phi_F$ is constructed such that the product $\phi_F F$ is free of kinematic singularities on the unit disk $|z|<1$ \cite{Boyd:1994tt,Boyd:1997qw,Caprini:1997mu,Becher:2005bg,Arnesen:2005ez}. The product of outer function and form factor is then commonly expressed as a power series in $z$,
which is bounded in the semileptonic region.
Powers of $z$ are orthonormal with respect to the scalar product
\begin{equation}
    \braket{z^n|z^m}
        \equiv  \oint_{|z| = 1} \frac{\text{d}z}{iz} \,z^{n,*} z^m
        =       \int_{-\pi}^{+\pi} \text{d}\alpha \, z^{n,*} z^m\big|_{z=e^{i\alpha}}
        =       2\pi \delta_{nm}\,,
\end{equation}
that is, when integrated over the entire unit circle.
As a consequence, for an analytic function on the $z$ unit disk that is square-integrable on the $z$ unit circle,
the Fourier coefficients exist only for positive index $n$ and coincide with the Taylor coefficients for an
expansion in $z=0$. The contribution to the dispersive bound can then be expressed as the 2-norm of the Taylor coefficients.
For more details of the derivation, we refer the reader to Ref.~\cite{Caprini:2019osi}. \\

For $b\to s$ transitions, $\bar{B}_s\pi$ intermediate states produce the lowest-lying branch cut.
However, production of a $\bar{B}_s\pi$ state from the vacuum through a $\bar{s}b$ current violates isospin symmetry
and is therefore strongly suppressed, and is forbidden in lattice-QCD calculations with $m_u=m_d$. The production of $\bar{B}_s\pi\pi$ states is allowed by isospin symmetry, but the matrix elements for this process are still expected to be small due to the three-particle structure. For the purpose of this analysis we set $t_+$ to the numerically most relevant branch point, {\it i.e.}, to\footnote{Note that our method is still equally applicable for other, lower choices of $t_+$.}
\begin{equation}
    t_+ \equiv (m_{B} + m_{K})^2 \, .
\end{equation}
The integral contribution for $\bar{B}K$ intermediate states can then be mapped onto the entire unit circle in $z$ as discussed above, and their contributions to the dispersive bound can be expressed as the 2-norm of their Taylor coefficients.  However, intermediate states with larger pair-production thresholds cover only successively smaller \emph{arcs of the
unit circle}, and the correspondence of the 2-norm of the Taylor coefficients and their contributions to the
dispersive bound does not hold any longer.
The branch point at $t_+$ arises from scattering into on-shell $\bar{B}K$ intermediate states.

In the following, we discuss the application of the series expansion to baryon-to-baryon form factors in the presence
of a dispersive bound.
The main difference between our approach and other parametrizations is that we do not assume the lowest branch point $t_+$ to coincide
with the baryon/antibaryon threshold $t_\text{th} > t_+$. As a consequence, the contour integral representing the form factor's
contribution to its bound is supported only on the arc of the unit circle with opening angle $2 \alpha_{\Lambda_b\Lambda}$, where
\begin{equation}
    \alpha_{\Lambda_b \Lambda} = \arg z((\mLam + \mLamB)^2)\,.
    \label{eq:LbL-angle}
\end{equation}
Specifically, Eq.~(\ref{eq:th:parametrization:bound-V1}) becomes
\begin{align}
    \nonumber
        1  &\geq \frac{1}{48\pi^2 \chi_V^{J=1}(Q^2)\big|_\text{OPE}} \int_{-\alpha_{\Lambda_b \Lambda}}^{+\alpha_{\Lambda_b \Lambda}} \text{d}\alpha  \left |\frac{\text{d}z(\alpha)}{\text{d}\alpha} \frac{\text{d}t(z)}{\text{d}z} \right| \frac{\sqrt{\lambda(\mLamB^2,\mLam^2,t) }}{t^2 (t-Q^2)^{n+1}}  s_{-}(t)  \left( (\mLamB +\mLam)^2 |f_{0}^V(t)|^2 +2 t |f_{\perp}^V(t)|^2  \right)  \\
    \label{eq:th:parametrization:bound-analytic}
        & \equiv  \int_{-\alpha_{\Lambda_b \Lambda}}^{+\alpha_{\Lambda_b \Lambda}} \text{d}\alpha \, \left(|\phi_{f_0^V}(z)|^2  |f_{0}^V(z)|^2 +|\phi_{f_\perp^V}(z)|^2  |f_{\perp}^V(z)|^2 \right)_{z=e^{i\alpha}}\, ,
\end{align}
where $t = t(z(\alpha))$, and we dropped the one-particle contributions for legibility. Here, $\phi_{f_0^V}(z),\phi_{f_\perp^V}(z)$ are the outer functions for the form factors $f_0^V$ and $f_\perp^V$. The full list of expressions for the outer functions of all baryon-to-baryon form factors is compiled in Appendix~\ref{app:outerfunction}.

A form factor's contribution to the bound is expressed in terms of an integral with a positive definite integrand.
Hence, we immediately find that a parametrization that assumes integration over the full unit circle rather than the
relevant pair production arc $|\alpha| < \alpha_{\Lambda_b\Lambda}$
\emph{overestimates} the saturation of the dispersive bound due to that form factor.
To express the level of saturation due to each term in \refeq{th:parametrization:bound-analytic}
as a 2-norm of some coefficient sequence, we
expand the form factors in a basis of polynomials $p_n(z)$ \cite{Gubernari:2020eft}.
These polynomials must be orthonormal with respect to the scalar product
\begin{equation}
    \braket{p_n|p_m}
        \equiv  \oint_{\substack{|z| = 1\\ |\arg z| \leq \alpha_{\Lambda_b \Lambda}}} \frac{\text{d}z}{iz} \, p^*_n(z)\, p_m(z)
        =       \int_{-\alpha_{\Lambda_b \Lambda}}^{+\alpha_{\Lambda_b \Lambda}} \text{d}\alpha \, p^*_n(z)\, p_m(z)\big|_{z=e^{i\alpha}}
        =       \delta_{nm} \,.
    \label{eq:szego-ortho}
\end{equation}
The polynomials $p_n(z)$ are the Szeg\H{o} polynomials~\cite{Simon2004OrthogonalPO}, which can be derived via the the Gram-Schmidt procedure;
see details in Appendix~\ref{app:gram-schmidt}.
A computationally efficient and numerically stable evaluation of the polynomials
can be achieved using the Szeg\H{o} recurrence relation~\cite{Simon2004OrthogonalPO}, which we use in the reference implementation
of our parametrization as part of the \EOS software. The first five so-called Verblunsky coefficients that uniquely generate the polynomials
are listed in Appendix~\ref{app:gram-schmidt}.
\\

Truncating the series at order $N$, our parametrization of the local form factors now takes the form
\begin{align}
    \label{eq:para}
    f_\lambda^\Gamma(q^2)\bigg|_{N}
        & = \frac{1}{\mathcal{P}(q^2) \, \phi_{f_\lambda^\Gamma}(z)}
            \sum_{i=0}^{N} a^{i}_{f_\lambda^\Gamma}\big|_{N} \, \, p_{i}(z),
\end{align}
where $\mathcal{P}(q^2) = z(q^2; t_0 = m^2_{\text{pole}}, t_+)$ is the Blaschke factor, $\phi_{f_\lambda^\Gamma}(z)$ is the outer function, and $p_{i}(z)$ are the orthonormal polynomials.
The Blaschke factor takes into account bound-state poles below the lowest branch point $t_+$ without changing the
contribution to the dispersive bound~\cite{Caprini:2019osi}.
Here, we assume each form factor to have a single bound-state pole, with the masses given in Table~\ref{tab::mpoles}.
For our parametrization, we choose $t_0 = q^2_{\text{max}} = (\mLamB - \mLam)^2$.
Our choice of $t_0$ means that the entire semileptonic phase space is mapped onto an interval of the \emph{positive} real $z$ axis.
Our parametrization features most of the benefits inherent to the BGL parametrization for meson-to-meson form factors~\cite{Boyd:1994tt}, with one exception. 
The BGL parametrization uses the $z^n$ monomials, which are bounded on the open unit disk.
As a consequence, the form-factor parametrization for processes such as $\bar{B}\to D$ are an \emph{absolutely convergent} series~\cite{Caprini:2019osi}.
This benefit does not translate to the baryon-to-baryon form factors.\footnote{%
    It also does not transfer to form factors for processes such as $\bar{B}_s\to D_s$ or $\bar{B}_s\to \bar{K}$,
    which suffer from the same problem: branch cuts below their respective pair-production thresholds. Our approach can be adjusted
    for these form factors.
}
The polynomials $p_n$ are not bounded
on the open unit disk. In fact, the Szeg\H{o} recurrence relation combined with the Szeg\H{o} condition
provide that $p_n(z = 0)$ increase exponentially with $n$ for large $n$.
Nevertheless, our proposed approach provides a benefit over a parametrization in terms of the $z$ monomials in absence
of any bound on their coefficients.
In appendix \ref{app:convergence}, we show that this growth does not spoil the convergence of the parametrization, and we provide an estimate for an upper bound on the truncation error.

Based on Eqs.~(\ref{eq::vector-FF})--(\ref{eq::tensor-FF}), we arrive at \emph{strong unitarity bounds} on the form-factor coefficients: 
\begin{align}
    \label{eq:sUB-1}
        \sum_{i=0}^N |a_{f_t^V}^i|^2 &\leq 1 - \frac{\chi_V^{J=0}\big|_\text{1pt\space\space}}{\chi_V^{J=0}\big|_\text{OPE}}\, , & \sum_{i=0}^N |a_{f_t^A}^i|^2 \leq 1 - \frac{\chi_A^{J=0}\big|_\text{1pt\space\space}}{\chi_A^{J=0}\big|_\text{OPE}} \, , \\
    \label{eq:sUB-2}
        \sum_{i=0}^N \left \{|a_{f_0^V}^i|^2 +|a_{f_\perp^V}^i|^2 \right \}  &\leq 1 - \frac{\chi_V^{J=1}\big|_\text{1pt\space\space}}{\chi_V^{J=1}\big|_\text{OPE}} \, , & \sum_{i=0}^N \left\{ |a_{f_0^A}^i|^2 +|a_{f_\perp^A}^i|^2 \right\} \leq 1 - \frac{\chi_A^{J=1}\big|_\text{1pt\space\space}}{\chi_A^{J=1}\big|_\text{OPE}} \, , \\
    \label{eq:sUB-3}
        \sum_{i=0}^N  \left \{|a_{f_0^T}^i|^2 +|a_{f_\perp^T}^i|^2  \right \} &\leq 1 - \frac{\chi_T^{J=1}\big|_\text{1pt\space\space}}{\chi_T^{J=1}\big|_\text{OPE}} \, , & \sum_{i=0}^N \left\{ |a_{f_0^{T5}}^i|^2 +|a_{f_\perp^{T5}}^i|^2 \right\}  \leq 1 - \frac{\chi_{T5}^{J=1}\big|_\text{1pt\space\space}}{\chi_{T5}^{J=1}\big|_\text{OPE}}\, .
\end{align}
Note that here we also subtracted the one-particle contributions, which are discussed in Sec.~\ref{sec:one-particle}.
However, this subtraction decreases the bound by only $\sim 10 \%$. In our statistical analysis
of only $\Lambda_b\to \Lambda$ form factors, we find that this subtraction is not yet numerically significant.
Nevertheless, we advocate to include the one-particle contributions in global fits of the
known local $b\to s$ form factors, where their impact will likely be numerically relevant.

\begin{table}[t]
\begin{center}
\renewcommand{\arraystretch}{1.25}
\begin{tabular}{C{4cm} C{3cm} C{3cm}}
    \toprule
    Form factor & Pole spin-parity $J^P$ & $m_{\text{pole}}$ in GeV \\
    \midrule
    \\[-1em]
    $f_0^V, f_\perp^V, f_0^T, f_\perp^T$        & $1^-$    & 5.416 \\
    $f_t^V$                                     & $0^+$    & 5.711 \\
    $f_0^A, f_\perp^A, f_0^{T5}, f_\perp^{T5}$  & $1^+$    & 5.750 \\
    $f_t^A$                                     & $0^-$    & 5.367 \\
    \bottomrule
\end{tabular}
\renewcommand{\arraystretch}{1.0}
\end{center}
\caption{List of $B_s$ meson pole masses appearing in the different form factors. The values are taken from Refs.~\cite{Lang:2015hza,PDG2020}.}
\label{tab::mpoles}
\end{table}

At this point, we have not yet employed the endpoint relations given in Eqs.~(\ref{eq:ep1}) - (\ref{eq:ep3}). By using the endpoint relations, we can express the zeroth coefficient of $f_t^V, f_t^A, f_\perp^A, f_\perp^T, f_0^{T5}$ in terms of coefficients of other form factors.

Our proposed parametrization has two tangible benefits. First, each form-factor parameter $a_k$ is bounded in magnitude,
$|a_k| \leq 1$.
The $N$ dimensional parameter space is therefore restricted to the hypercube $[-1, +1]^N$.
We refer to this type of parameter bound as the \emph{weak bound}\footnote{%
    Our definitions of \emph{weak} and \emph{strong} bounds differ from the definitions proposed
    in Ref.~\cite{Bigi:2017jbd}. There, what we call the \emph{weak bound} is not considered
    in isolation, and what we call the \emph{strong bound} is labelled a ``weak bound'', in contrast
    to a ``strong bound'' that affects more than one decay process.
}.
It facilitates fits to theoretical or phenomenological inputs on the form factors, since the choice
of a prior is not subjective.
Second, the form-factor parameters are restricted by the \emph{strong bounds} \refeq{sUB-1} to \refeq{sUB-3}.
In the absence of the small number of exact relations between the form factors that we discussed earlier,
this strong bound is in fact an upper bound on the sum of the squares of the form-factor parameters.
As a consequence, the parameter space is further restricted to the combination of four hyperspheres,
one per bound.\footnote{%
    The form factor relations mix the parameters of form factors that belong to different
    strong bounds, thereby making a geometric interpretation less intuitive.
}
The strong bounds imply that the sequence of form-factor parameters asymptotically falls off faster
than $1/\sqrt{k}$. This behaviour does not prove absolute convergence of the series expansion of the form factors,
which would require a fall off that compensates the exponential growth of the polynomials.
Nevertheless, we argue in Appendix \ref{app:convergence} that the series converges rapidly.
Below, we check empirically if the strong bound suffices to provide bounded uncertainties
for the form factors in truncated expansions.

\section{Statistical Analysis}
\label{sec::Numerical-Analysis}

\subsection{Data Sets}

To illustrate the power of our proposed parametrization, we carry out a number of Bayesian analyses
of the lattice QCD results for the full set of $\Lambda_b\to \Lambda$ form factors as provided in Ref.~\cite{Detmold:2016pkz}.
These analyses are all carried out using the \EOS software~\cite{vanDyk:2021sup}, which has been modified
for this purpose. Our proposed parametrization for the $\Lambda_b\to\Lambda$ form factors is implemented as of
\EOS version 1.0.2~\cite{EOS:v1.0.2}.
The form factors are constrained by a multivariate Gaussian likelihood that jointly describes
synthetic data points of the form factors in the continuum limit and at physical quark masses, up to three per form factor. Each data point
is generated for one of three possible values of the momentum transfer $q^2$: $q^2_i \in \{13, 16, 19 \}$ GeV$^2$.
The overall $q^2$ range is chosen based on the availability of lattice QCD data points in Ref.~\cite{Detmold:2016pkz}.
The synthetic data points are illustrated by black crosses in Figs.~\ref{fig:formfactor-nominal}--\ref{fig:formfactor-truncation}.

Reference~\cite{Detmold:2016pkz} provides two sets of parametrizations of the form factors in the continuum limit
and for physical quark masses, obtained from one ``nominal'' and one ``higher-order'' fit to the lattice data.
The nominal fit uses first-order $z$ expansions, which are modified with correction terms that describe the
dependence on the lattice spacing and quark masses. The higher-order fit uses second-order $z$ expansions
and also includes higher-order lattice-spacing and quark-mass corrections.
The parameters that only appear in the higher-order fit are additionally constrained with Gaussian priors.
In the case of lattice spacing and quark masses, these priors are well motivated by effective field theory
considerations~\cite{Detmold:2016pkz}.
In the higher-order fit, the coefficients $a^2_{f_\lambda^\Gamma}$
of the $z$ expansion are also constrained with Gaussian priors, centered around zero and widths equal to twice the magnitude
for the corresponding coefficients $a^1_{f_\lambda^\Gamma}$ obtained within the nominal fit.
This choice of prior was less well motivated but has little effect in the high-$q^2$ region.
Ref.~\cite{Detmold:2016pkz} recommends to use the following procedure for evaluating the form factors in phenomenological applications:
the nominal-fit results should be used to evaluate the central values and statistical uncertainties,
while a combination of the higher-order-fit and nominal-fit results should be used to estimate systematic uncertainties as explained in Eqs.~(50)--(56) in Ref.~\cite{Detmold:2016pkz}.

To generate the synthetic data points for the present work, we first updated both the nominal and the higher-order fits of Ref.~\cite{Detmold:2016pkz} with
minor modifications: we now enforce the endpoint relations
among the form factors at $q^2=0$ \emph{exactly}, rather than approximately as done in Ref.~\cite{Detmold:2016pkz}, and
we include one additional endpoint relation $f_\perp^{T5}(0)=f_\perp^T(0)$, which is not used in Ref.~\cite{Detmold:2016pkz}.

The synthetic data points for $f_0^V$, $f_0^A$ and $f_\perp^T$ at $q^2 = 13 \,\,{\rm GeV}^2$, and $f_0^A$ and $f_0^{T5}$ at 19\,\,GeV$^{2}$,
have strong correlation with other data points. This can be understood, since five exact relations hold for these form factors
either at $q^2 = 0$ or $q^2 = (\mLamB - \mLam)^2$ between pairs of form factors. We remove the synthetic data points
listed above, which renders the covariance matrix regular and positive definite. We arrive at a 25-dimensional multivariate
Gaussian likelihood. The likelihood is accessible under the name
\begin{center}
    \texttt{Lambda\_b->Lambda::f\_time+long+perp\^{}V+A+T+T5[nominal,no-prior]\@DM:2016A}
\end{center}
as part of the constraints available within the \EOS software.

\subsection{Models}

In this analysis, we consider a variety of statistical models.
First, we truncate the series shown in Eq.~(\ref{eq:para}) at $N=2$, 3 or 4. 
The number of form-factor parameters is $10 (N + 1)$, due to a total
of ten form factors under consideration. 
Since we implement the five form factors relations \emph{exactly},
the number of fit parameters is smaller than the number of form-factor parameters by five.
Hence, we arrive at between $P=25$ and $P=45$ fit parameters.
We use three different types of priors in our analyses.
An analysis labelled ``w/o bound'' uses a uniform prior, which is chosen to contain at least $99\%$ of the integrated
posterior probability. An analysis labelled ``w/ weak bound'' uses a uniform prior on the hypercube $[-1, +1]^P$,
thereby applying the weak bound for all fit parameters.
An analysis labelled ``w/ strong bound'' uses the same prior as the weak bound. In addition, we modify the posterior
to include the following element, which can be interpreted either as an informative non-linear prior or a
factor of the likelihood.
For each of the six bounds $B(\lbrace a_n\rbrace)$, we add the penalty term~\cite{Bordone:2019vic}
\begin{equation}
    \begin{cases}
        0                 & \rho_B < 1,\\
        100 (\rho_B - 1)^2 & \text{otherwise}
    \end{cases}\,
\end{equation}
to $-2 \ln \text{Posterior}$.
Here, $\rho_B = \sum_n |a_n|^2$, and the sum includes only the parameters affected by the given bound $B$.
The additional terms penalize parameter points that violate any of the bounds with a one-sided $\chi^2$-like term.
The factor of $100$ corresponds to the inverse square of the relative theory uncertainty on the bound,
which we assume to be $10\%$.
This uncertainty is compatible with the results obtained in Ref.~\cite{Bharucha:2010im}.
In the above, we use unity as the largest allowed saturation of each bound. As discussed in Sec.~\ref{sec:th:had-repr},
one-body and mesonic two-body contributions to the bounds are known. They could be subtracted from the upper bounds.
However, we suggest here to include these contributions on the left-hand side of the bound in a global analysis
of the available $b\to s$ form factor data. A global analysis clearly benefits from this treatment, which induces
non-trivial theory correlations among the form-factor parameters across different processes.
It also clearly goes beyond the scope of the present work.

For $N=2$, the number of parameters is equal to the number of data points, and we arrive at zero degrees of freedom.
For $N > 2$, the number of parameters exceeds the number of data points.
Hence, a frequentist statistical interpretation is not possible in these cases.
Within our analyses, we instead explore whether the weak or strong bounds suffice to
limit the {\it a posteriori} uncertainty on the form factors, despite having zero or negative degrees of freedom.\\

\subsection{Results} 

\begin{figure}[p!]
    \begin{tabular}{cc}
        \includegraphics[width=.4\textwidth]{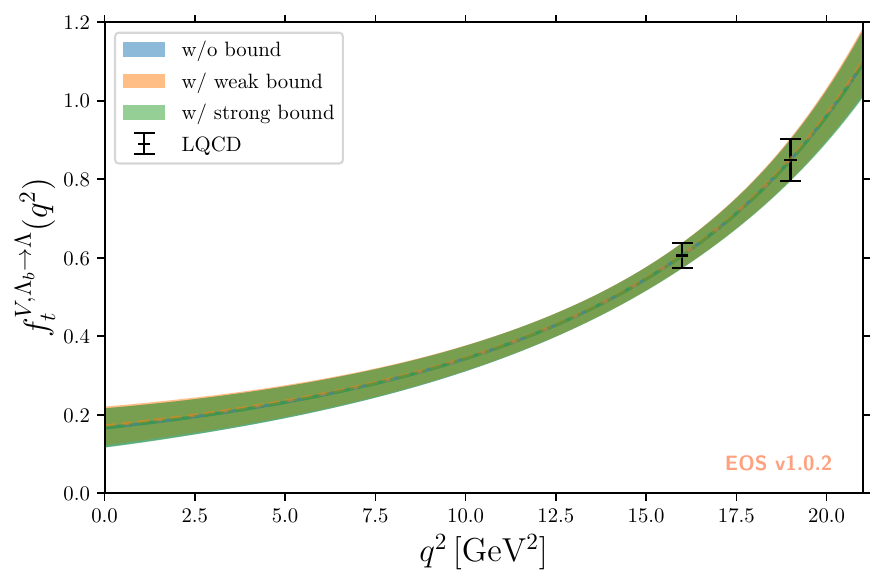}             &
        \includegraphics[width=.4\textwidth]{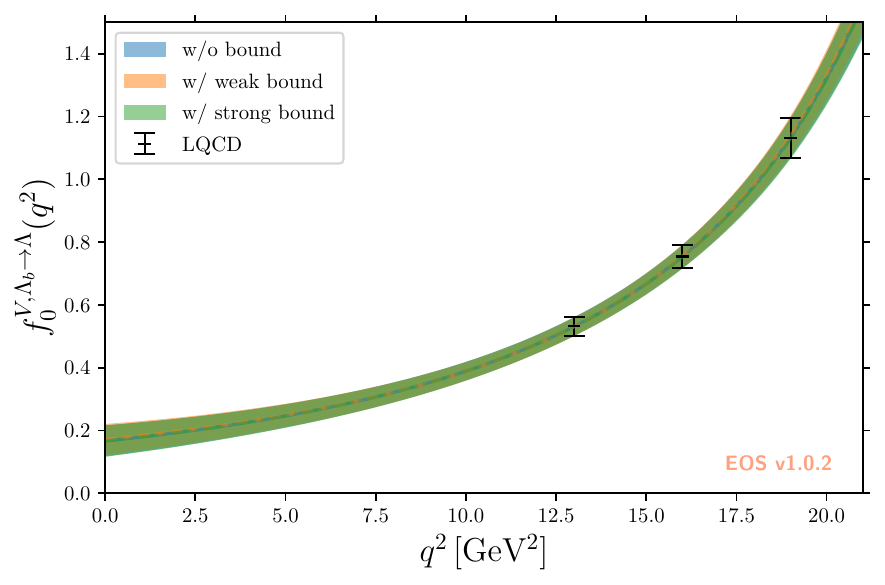} \\[-.5em]
        \includegraphics[width=.4\textwidth]{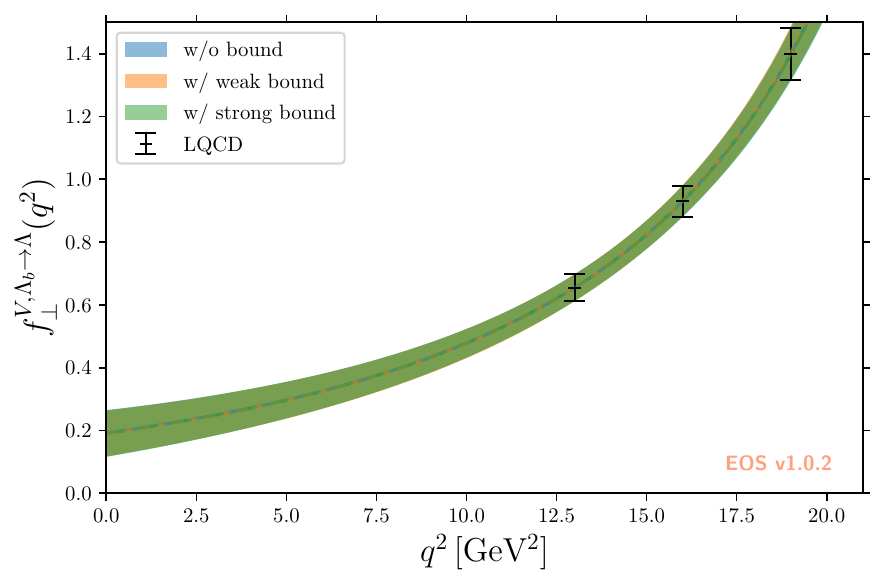}             &
        \includegraphics[width=.4\textwidth]{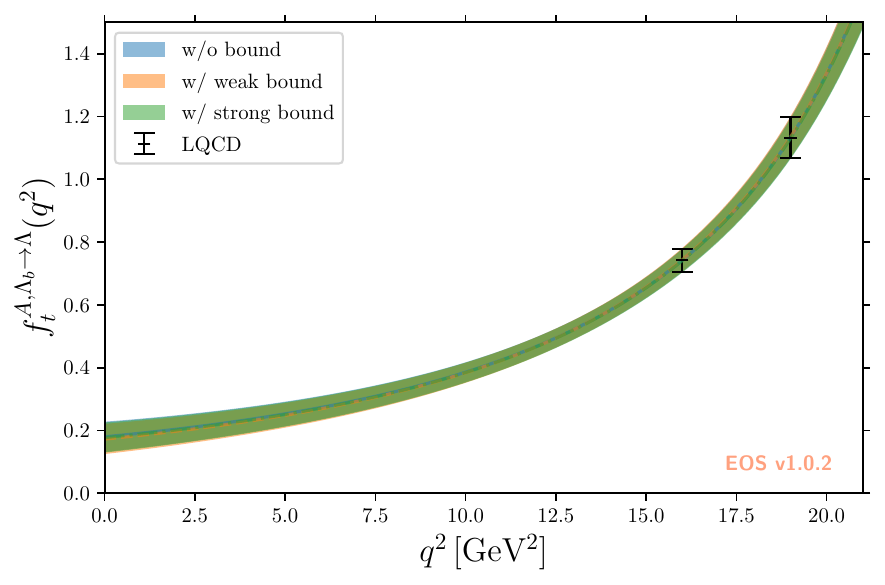} \\[-.5em]
        \includegraphics[width=.4\textwidth]{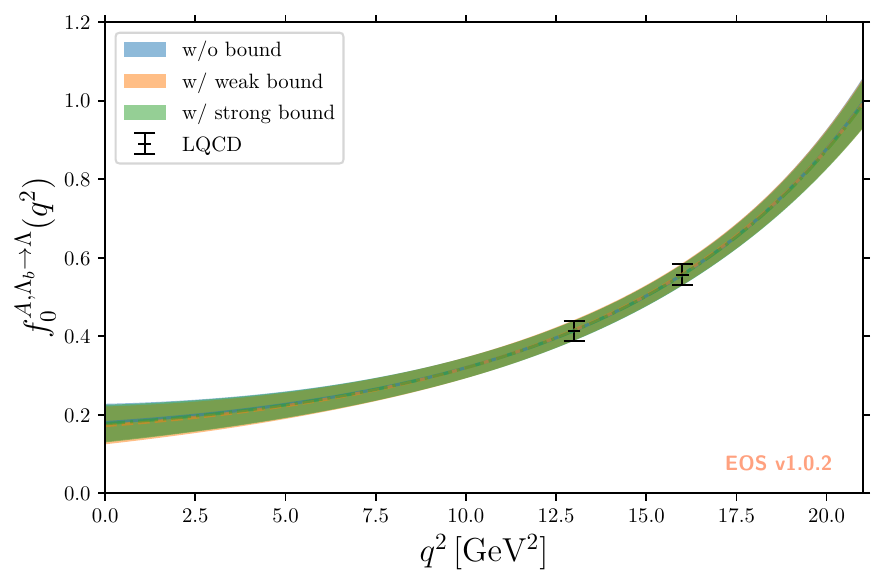}             &
        \includegraphics[width=.4\textwidth]{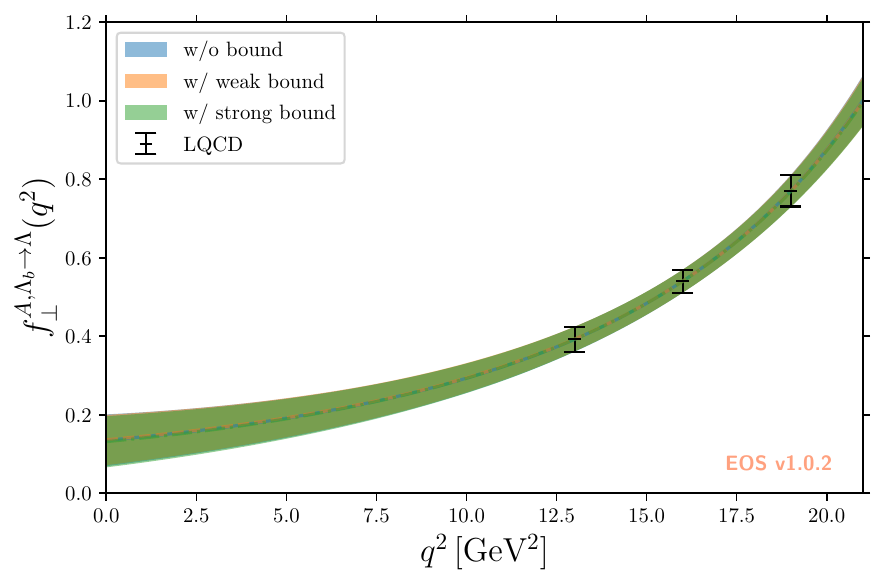} \\[-.5em]
        \includegraphics[width=.4\textwidth]{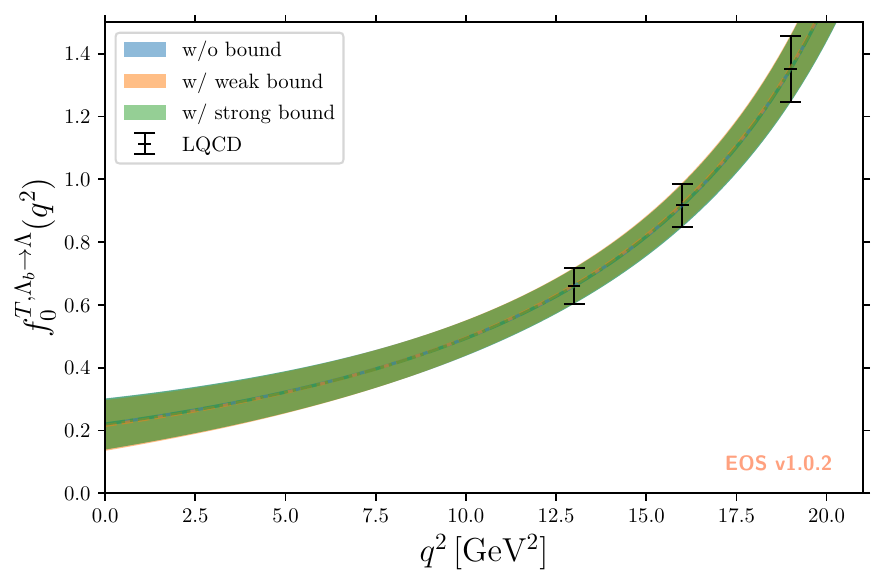}             &
        \includegraphics[width=.4\textwidth]{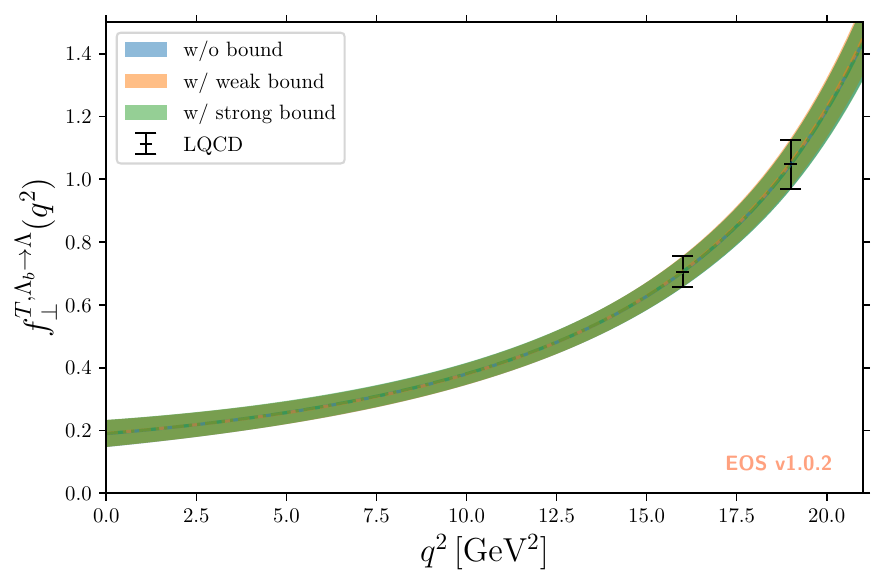} \\[-.5em]
        \includegraphics[width=.4\textwidth]{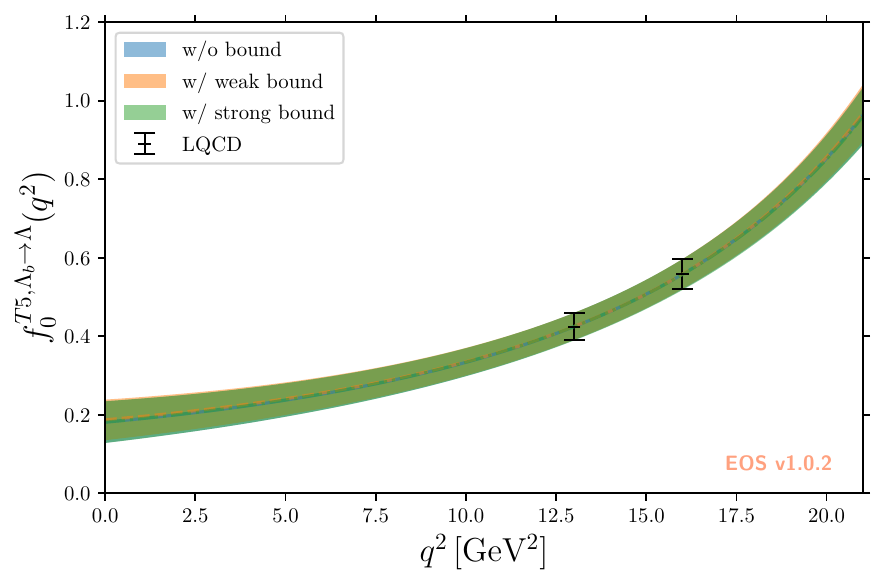}            &
        \includegraphics[width=.4\textwidth]{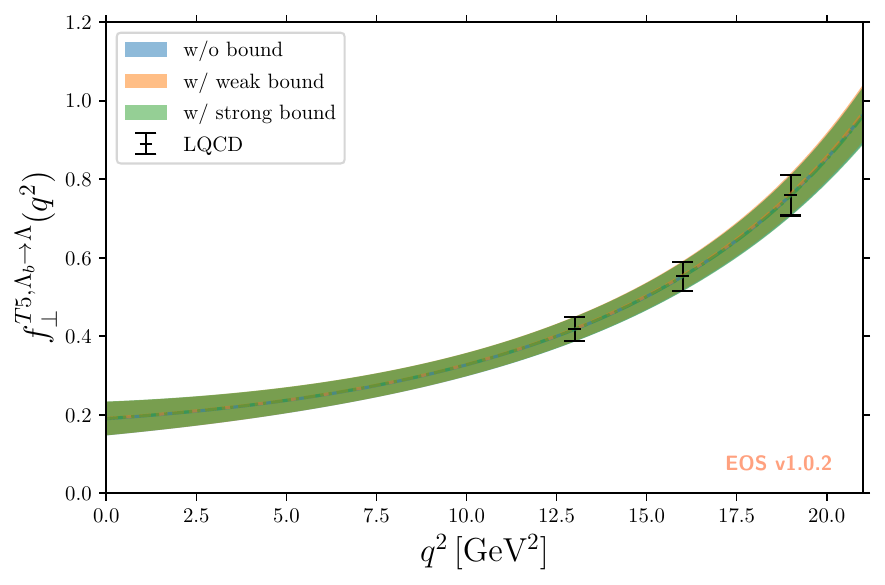}  
    \end{tabular}
    \caption{%
        Uncertainty bands for the {\it a posteriori} form-factor predictions of the ten form factors.
        The bands comprise the central $68\%$ probability interval at every point in $q^2$.
        We show the form-factor results at $N =2$ in the absence of any bounds,
        using weak bounds $|a_{V,\lambda}^i| <1$, and using the strong bounds (see text), respectively. The markers indicate the synthetic lattice data points. 
    }
    \label{fig:formfactor-nominal}
    \vspace{-100pt}
\end{figure}

\begin{figure}[p]
    \begin{tabular}{cc}
        \includegraphics[width=.4\textwidth]{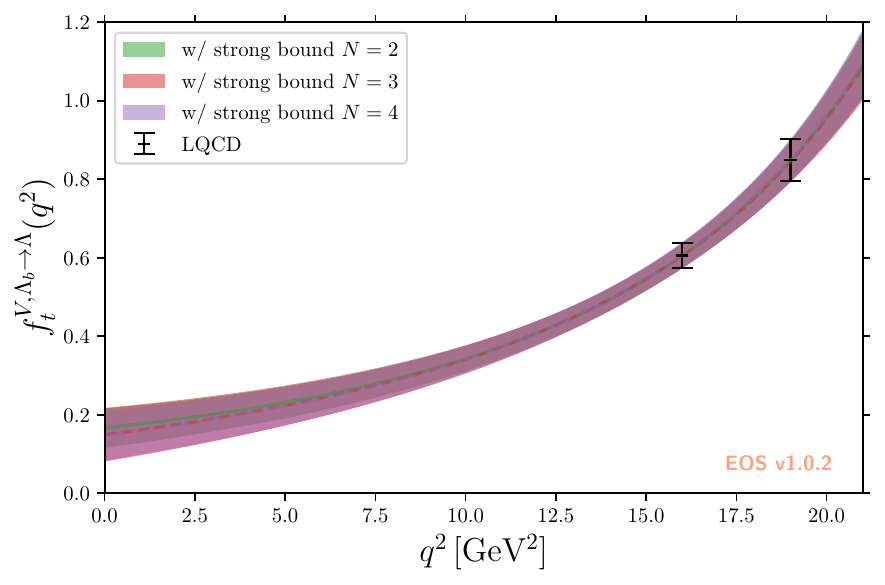}             &
        \includegraphics[width=.4\textwidth]{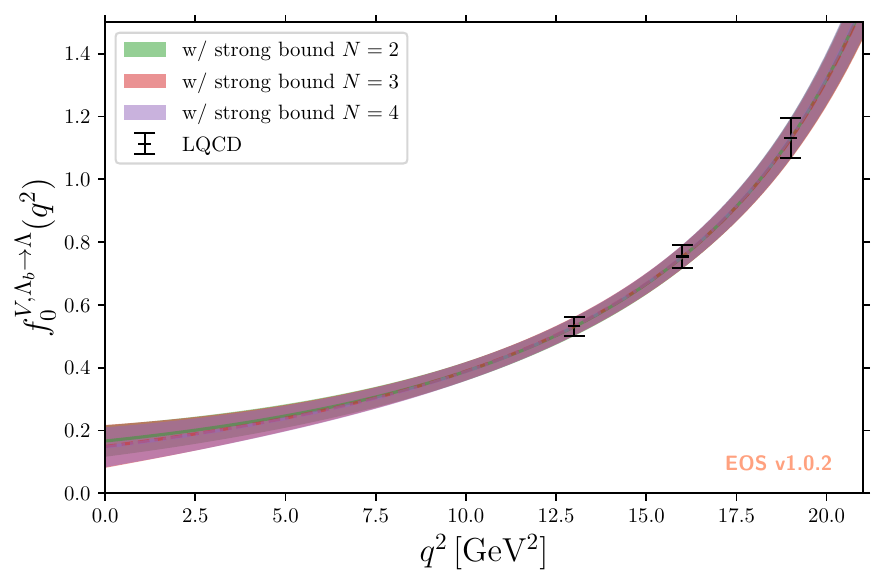} \\
        \includegraphics[width=.4\textwidth]{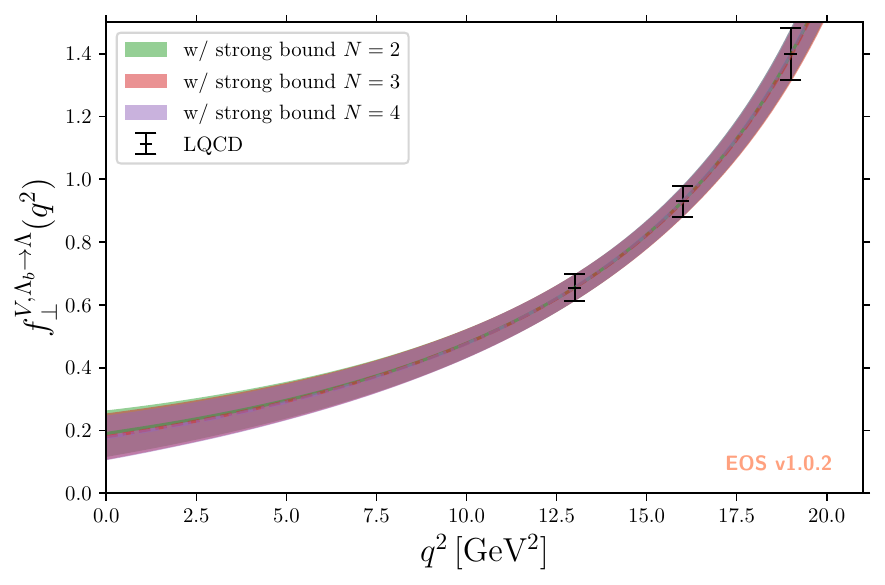}             &
        \includegraphics[width=.4\textwidth]{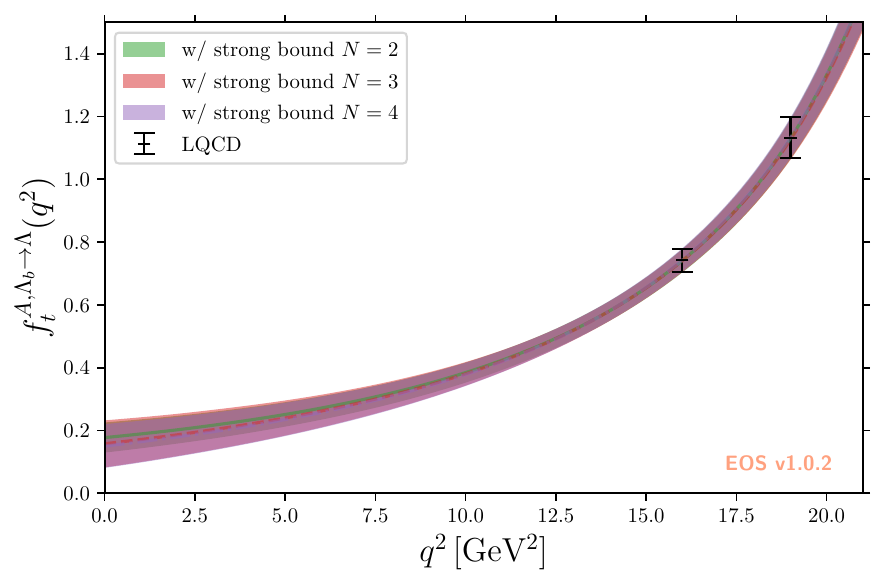} \\
        \includegraphics[width=.4\textwidth]{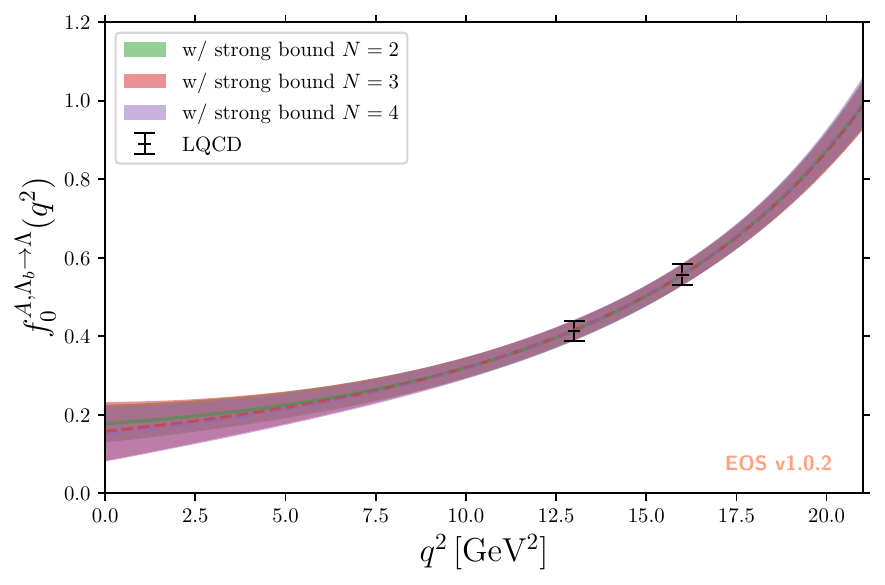}             &
        \includegraphics[width=.4\textwidth]{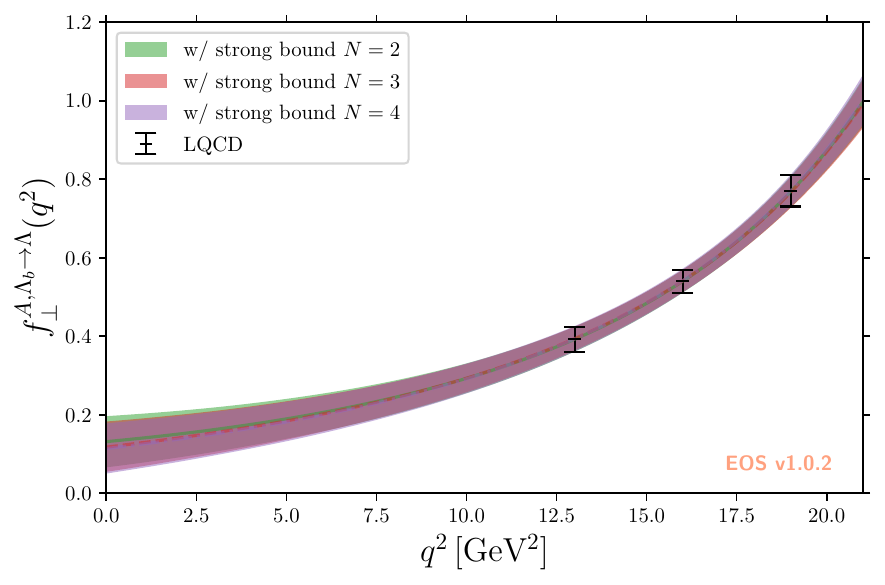} \\
        \includegraphics[width=.4\textwidth]{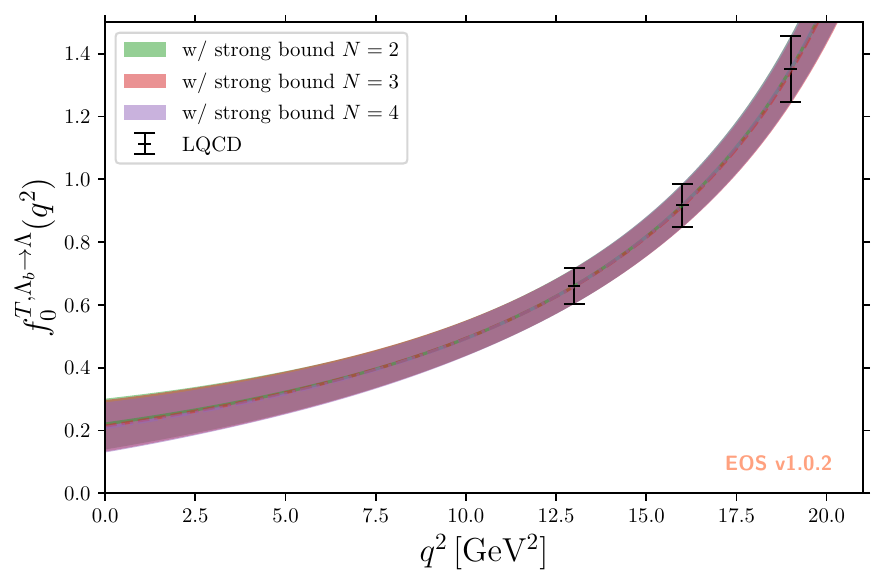}             &
        \includegraphics[width=.4\textwidth]{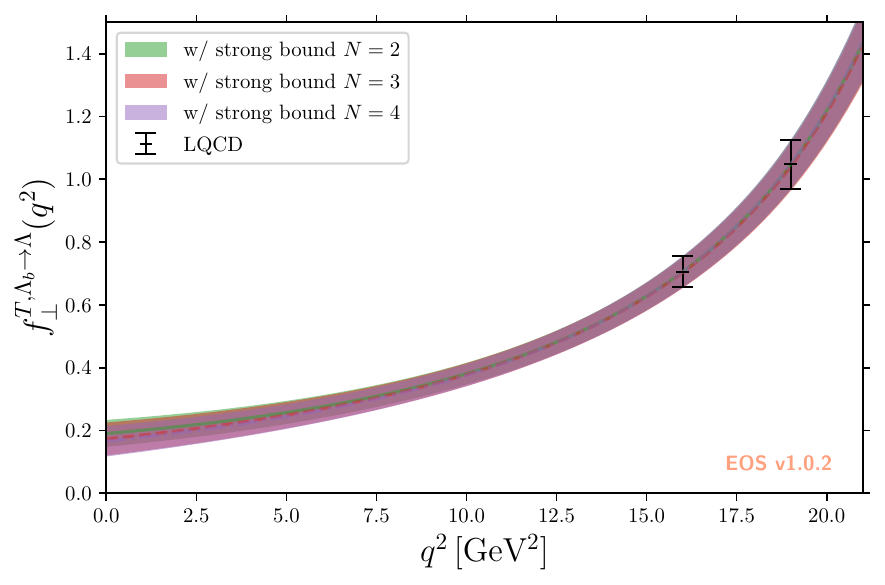} \\    
        \includegraphics[width=.4\textwidth]{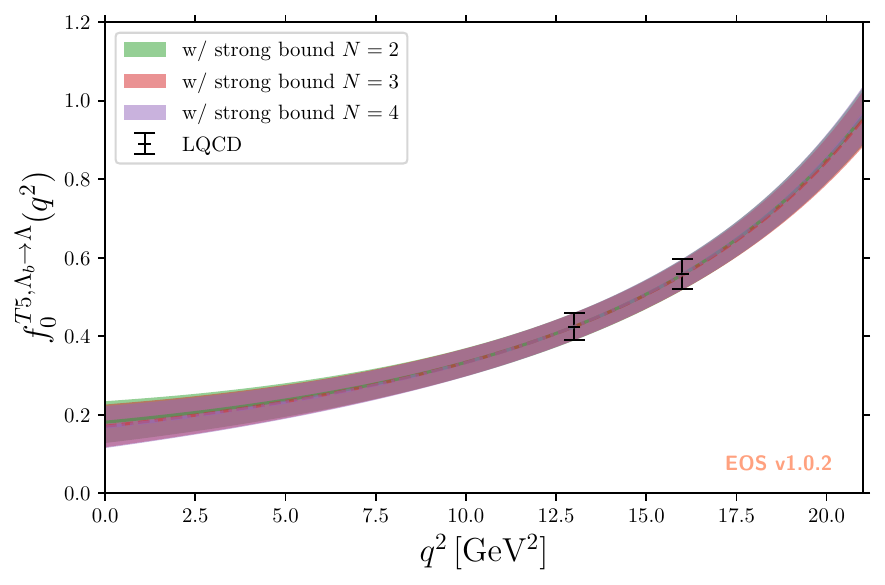}            &
        \includegraphics[width=.4\textwidth]{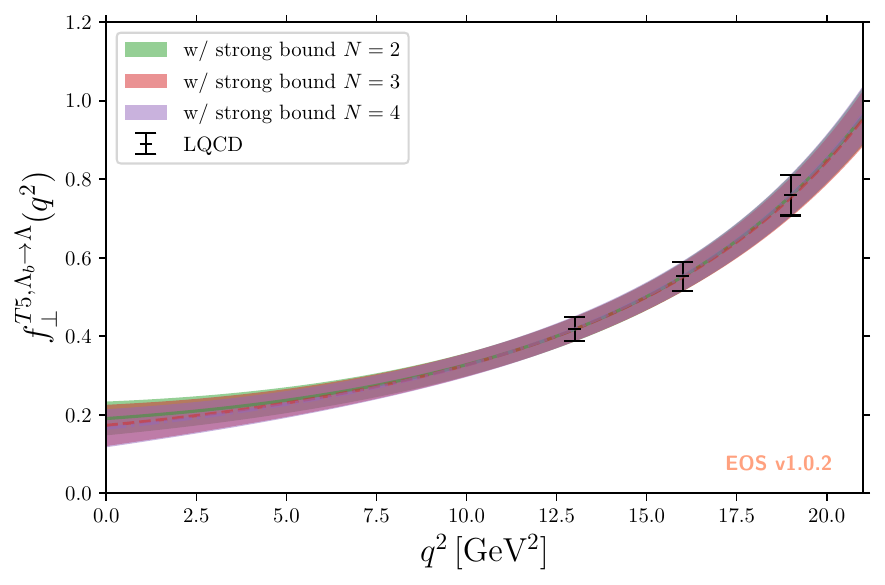}  
    \end{tabular}
    \caption{%
        Uncertainty bands for the {\it a posteriori} form-factor predictions of the ten form factors.
        The bands comprise the central $68\%$ probability interval at every point in $q^2$.
        We show the form-factor results at $N \in \{2, 3, 4\}$ when using the strong bound.
        Note that for $N > 2$ we have more parameters than data points. Finite uncertainty envelopes are
        enforced by the bound. The markers indicate the synthetic lattice data points. 
    }
    \label{fig:formfactor-truncation}
    \vspace{-100pt}
\end{figure}

We begin with three analyses at truncation $N=2$, using each of the three types of priors defined
above. In all three analyses, we arrive at the same best-fit point. This indicates clearly that the
best-fit point not only fulfills the weak bound, but also the strong bound. We explicitly confirm
this by predicting the saturation of the individual bounds at the best-fit point. These range
between $12\%$ (for the $1^-$ bound) and $33\%$ (for the $1^+$ bound), which renders the point
\emph{well within} the region allowed by the strong bound. Accounting for the known
one-particle contributions does not change this conclusion.
At the maximum-likelihood point, the $\chi^2$ value arising from the likelihood is compatible with zero
at a precision of $10^{-5}$ or better.
For each of the three analyses, we obtain a unimodal posterior and sample from the posterior using multiple Markov chains and the Metropolis-Hastings algorithm~\cite{Metropolis:1953am,Hastings:1970aa}. We use these samples to produce posterior-predictive distributions for each of the form factors,
which are shown in Figures \ref{fig:formfactor-nominal}--\ref{fig:formfactor-truncation} on the left-hand side. We observe that the strong bound
has some impact on the form factor uncertainties, chiefly far away from the region where synthetic
data points are available. For $N=2$, we do not find a significant reduction of the uncertainties due to the
application of the strong bound. Rather, it influences the shape of the form factors and suppresses the
appearance of local minima in the form factors close to $q^2 = 0$, which become visible when extrapolating to negative $q^2$.
The modified shape aligns
better with the naive expectation that the form factors rise monotonically with increasing $q^2$
below the first subthreshold pole. It also provides confidence that, with more precise lattice
QCD results, analyses of the nonlocal form factors at negative $q^2$ can be
undertaken. This opens the door toward analysis in the spirit of what has been proposed in Refs.~\cite{Bobeth:2017vxj,Gubernari:2020eft}.

\begin{figure}[t]
    \begin{tabular}{cc}
        \includegraphics[width=.49\textwidth]{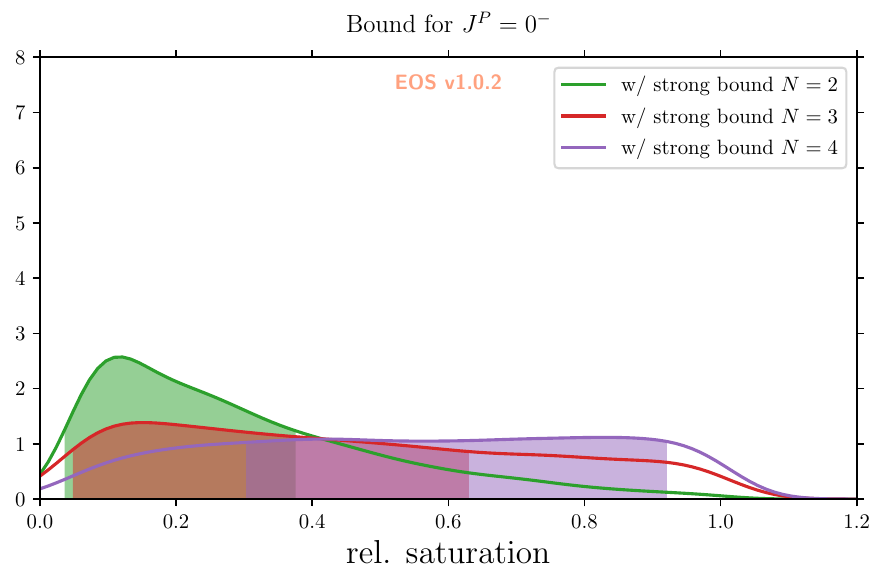}             &
        \includegraphics[width=.49\textwidth]{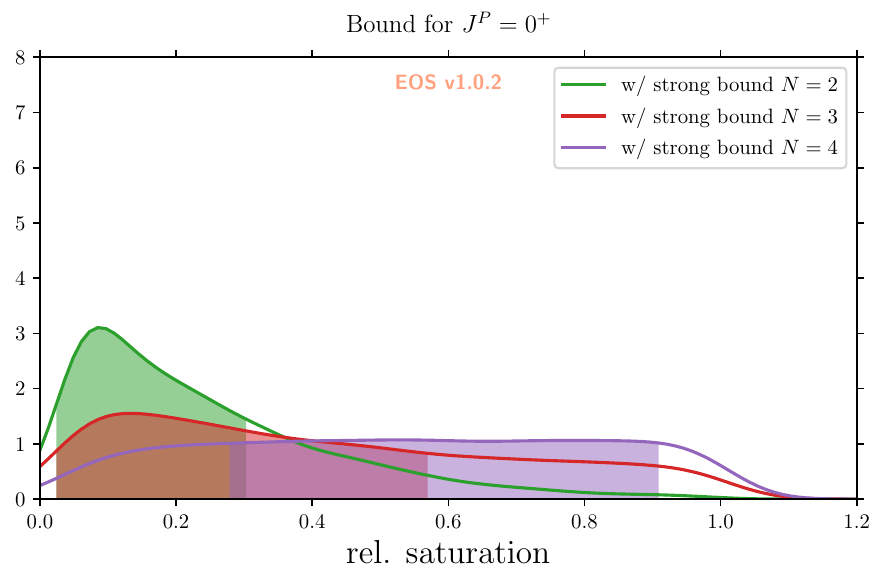} \\
        \includegraphics[width=.49\textwidth]{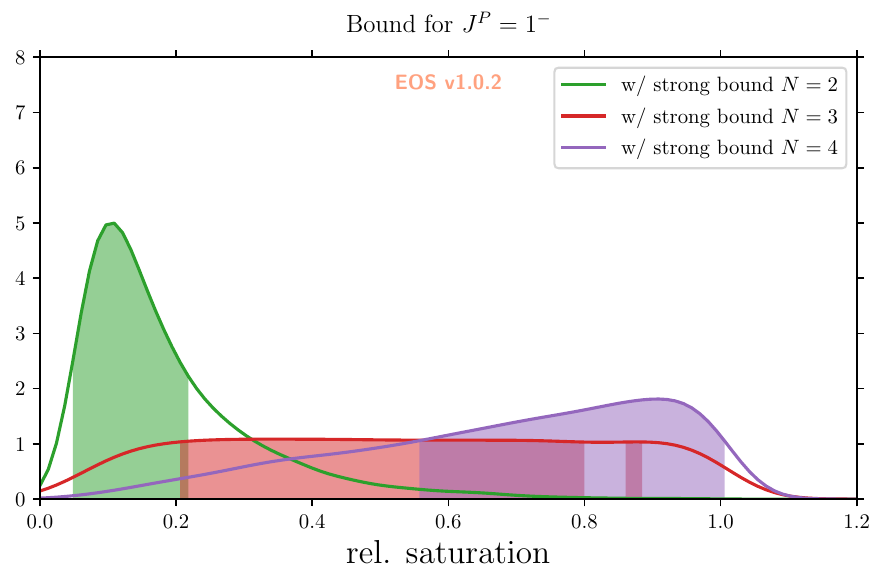}             &
        \includegraphics[width=.49\textwidth]{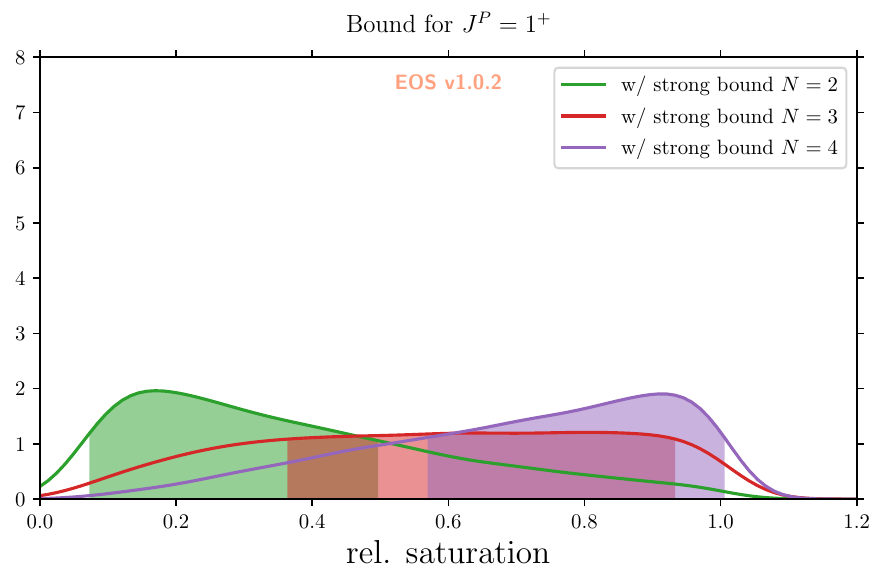} \\
        \includegraphics[width=.49\textwidth]{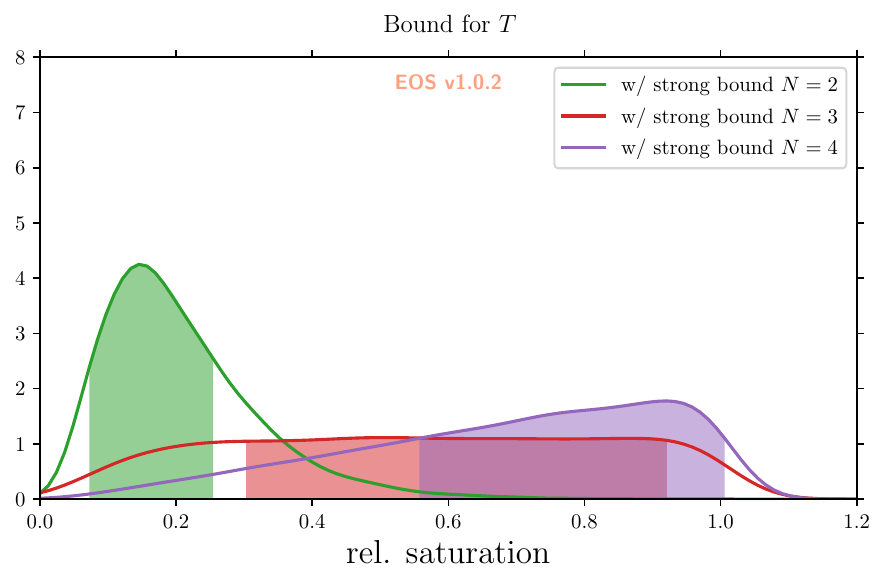}              &
        \includegraphics[width=.49\textwidth]{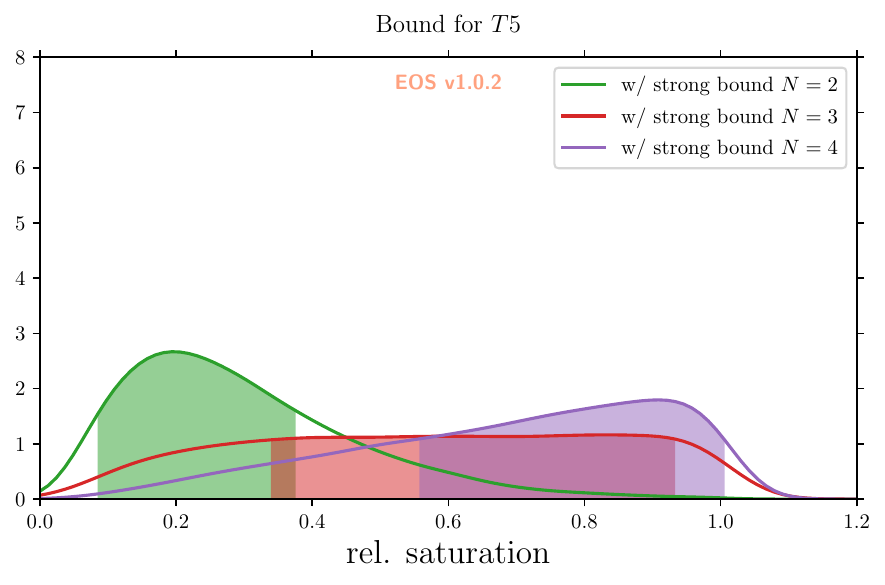}
    \end{tabular}
    \caption{%
        Relative saturation of the form factors with their respective spin-parity number $J^P$ obtained from posterior samples. The saturations are shown for different truncations of $N$, where the coefficients are constrained through the strong unitarity bound. 
        The vertical bands comprise the central $68\%$ probability interval.
    }
    \label{fig:saturation}
\end{figure}

We continue with three analyses using the strong bound, for $N=2$, $N=3$, and $N=4$. Note that, due to the nature of the orthonormal polynomials, in general any two sets of coefficients $\lbrace a_i\rbrace|_{N}$
and $\lbrace a_i\rbrace|_{N'}$ are not nested, i.e., the first $\min(N,N')$ elements of the sets are \emph{not}
identical (see Appendix \ref{app:convergence} for further discussion). Hence, the best-fit point for $N=2$ is not expected to be nested
within the $N=3$ and $N=4$ solutions, and the $N=3$ best-fit point is not nested within the $N=4$ solution.
In all three cases, we find a single point that maximizes the posterior.
For all three points we find that the bounds are fulfilled and consequently we obtain $\chi^2$ values consistent with zero.
The form-factor shapes are compatible between the $N=2$, $3$ and $4$ solutions.
We show the {\it a posteriori} form factor envelopes at 68\% probability together with the median values in
\reffig{formfactor-truncation}.
A clear advantage of our proposed parameterization is that the uncertainties in the large recoil region, {\it i.e.} away from the synthetic data points, do not increase dramatically when $N$ increases.
This is in stark contrast with a scenario without any bounds on the coefficients $a_n$,
where the {\it a posteriori} uncertainty for the form factors would be divergent for negative degrees of freedom.
This indicates that the bounds are able to constrain the parameterization even in an underconstrained analysis
and gives confidence that the series can be reliably truncated in practical applications of this method.
Figure~\ref{fig:saturation} shows the saturation of the strong bound for the different form factors with $N=2$, $3$ and $4$. 
For $N = 2$, the bounds are saturated between $10 - 30 \%$.
This is as large or even larger than the one-particle contributions, which saturate the bounds to $\sim 10\%$
and much larger than the two-particle mesonic contributions, which saturate the bounds by only $1$--$4\%$~\cite{Bharucha:2010im}.
As $N$ increases, the average saturation of the bounds increases. 
This is expected as additional parameters have to be included in the bound. The observed behaviour of the bound saturation provides further motivation for a global analysis of all
$b\to s$ form-factor data. 

Based on the updated analysis of the lattice data of  Ref.~\cite{Detmold:2016pkz},
we produce an {\it a posteriori} prediction for the tensor form factor $f_\perp^T$ at $q^2 = 0$
from our analyses.
We use this form factor as an example due to its phenomenological relevance in predictions of $\Lambda_b\to \Lambda\gamma$ observables.
Moreover, its location at $q^2 = 0$ provides the maximal distance between a phenomenologically relevant
quantity and the synthetic lattice QCD data points, thereby maximizing the parametrization's
systematic uncertainty. Applying the strong bound, we obtain
\begin{equation}
\begin{aligned}
    f_\perp^T(q^2 = 0)\big|_{N=2}
        & = 0.190 \pm 0.043\,, \\
    f_\perp^T(q^2 = 0)\big|_{N=3}
        & = 0.173 \pm 0.053\,, \\
    f_\perp^T(q^2 = 0)\big|_{N=4}
        & = 0.166 \pm 0.049\,. \\
\end{aligned}
\end{equation}
We observe a small downward trend in the central value and stable parametric uncertainties.
The individual bands are compatible with each other within their uncertainties.
We remind the reader that our results are obtained for negative degrees of freedom
and should therefore not be compared with the behaviour of a regular fit.
Our results should be compared with
\begin{equation}
    f_\perp^T(q^2 = 0)\big|_{\text{\cite{Detmold:2016pkz}}}
        = 0.166 \pm 0.072\,. \\
\end{equation}
This value and its uncertainty is obtained from the data and method described in Ref.~\cite{Detmold:2016pkz};
however, it includes the exact form-factor relation \refeq{ep3}, which has not been previously used.
Our parametrization exhibits a considerably smaller parametric uncertainty.

\section{Conclusion}
\label{sec:conc}

In this work we have introduced a new parametrization for the ten independent local $\Lambda_b \to \Lambda$ form factors.
Our parametrization has the advantage that the parameters are bounded, due to the use of orthonormal polynomials
that diagonalize the form factors' contribution within their respective dispersive bounds.
Using a Bayesian analysis of the available lattice QCD results for the $\Lambda_b \to \Lambda$ form factors, obtained in the continuum limit at physical quark masses, we illustrate that our parametrization
provides excellent control of systematic uncertainties when extrapolating from low to large hadronic recoil.
To that end, we investigate our parametrization for different truncations and observe that the extrapolation uncertainty does
not increase significantly within the kinematic phase space of $\Lambda_b\to\Lambda \ell^+\ell^-$ decays.
We point out that the dispersive bounds are able to constrain the form-factor uncertainties to such an extent that 
massively underconstrained analyses still exhibit stable uncertainty estimates.
This is a clear benefit compared to other parametrizations.

For future improvements of the proposed parametrization, one can insert the framework of dispersive bounds directly into the lattice-QCD analysis. Moreover, by including the one-particle contributions, as discussed in Sec.~\ref{sec:one-particle}, and other two-particle contributions, as discussed in Sec.~\ref{sec:two-particle}, in a global analysis of the available $b \to s$ form-factor data, we would expect even more precise results to be obtained for the form factors as the upper bound would be even more saturated.

\subsubsection*{Acknowledgements}

We thank Marzia Bordone, Nico Gubernari, Martin Jung, and M\'eril Reboud for helpful discussions.
We are grateful to Marvin Zanke for reporting two typos prior to publication.
The work of TB is supported by the Royal Society (UK). The work of SM is supported by the U.S. Department of Energy, Office of Science, Office of High Energy Physics under Award Number D{E-S}{C0}009913.
The work of MR is supported by the Deutsche Forschungsgemeinschaft (DFG, German Research Foundation) under grant 396021762 - TRR 257.
The work of DvD is supported by the DFG within the Emmy Noether Programme under grant DY-130/1-1
and the Sino-German Collaborative Research Center TRR110 ``Symmetries and the Emergence of Structure in QCD''
(DFG Project-ID 196253076, NSFC Grant No. 12070131001, TRR 110).
DvD was supported in the final phase of this work by the Munich Institute for Astro- and Particle Physics (MIAPP), which is
funded by the DFG under Germany’s Excellence Strategy – EXC-2094 – 390783311.

\clearpage 

\appendix

\section{Orthonormal polynomials}
\label{app:gram-schmidt}

In this section we discuss briefly how to obtain the orthonormal polynomials $p_n(z)$, which enter the series expansion in Eq. (\ref{eq:para}) to parameterize the form factors of the $\Lambda_b \to \Lambda$ transition. The functions can be derived with the Gram-Schmidt orthogonalization process in the basis $\{1, z, \dots, z^{N} \}$ and fulfill~\refeq{szego-ortho}.
The orthonormal functions are defined on the arc of the unit circle that covers the angle between $-\alpha_{\Lambda_b \Lambda}$ and $+\alpha_{\Lambda_b \Lambda}$, see Eq. (\ref{eq:LbL-angle}). The orthonormal polynomials are given by
\begin{align}
p_n(z) & = \frac{p^{\prime}_{n}(z) }{ \sqrt{\braket{p^{\prime}_{n}(z)| p^{\prime}_{n}(z) }}} \, ,    
\end{align}
where
\begin{align}
    p^{\prime}_{n}(z)
        & = z^n - \sum_{j=0}^{n-1} \frac{\braket{ p_j^{\prime}(z)| z^n}}{\braket{ p_j^{\prime}(z)| p_j^{\prime}(z)}} \cdot p_j^{\prime}(z)\,, &
    p'_0(z)
        & = 1
    \,.
\end{align}

The orthonormal polynomials for $\Lambda_b\to\Lambda$ can be evaluated efficiently using
the orthogonal Szeg\H{o} polynomials via a recurrence relation~\cite{Simon2004OrthogonalPO}. We use
\begin{equation}
\begin{aligned}
    \Phi_0(z)
        & = 1\,, &
    \Phi^*_0(z)
        & = 1\,, \\
    \Phi_n(z)
        & = z \Phi_{n - 1} - \rho_{n - 1} \Phi^*_{n - 1}\,, &
    \Phi^*_n(z)
        & = \Phi^*_{n - 1} - \rho_{n - 1} z \Phi_{n - 1}\,, &
\end{aligned}
\end{equation}
which holds for real $z$.
The orthonormal polynomials then follow from
\begin{equation}
\begin{aligned}
    p_n(z) & = \frac{\Phi_n(z)}{N_n}\,, &
    N_n    & = \left[2\alpha_{\Lambda_b\Lambda} \prod_{i=0}^{n-1} \left(1 - \rho_i^2\right)\right]^{1/2}\,,
\end{aligned}
\end{equation}
where $2\alpha_{\Lambda_b\Lambda} = 3.22198$ and the Verblunsky coefficients are
\begin{equation}
    \left\lbrace \rho_0, \dots \rho_4 \right\rbrace
    = \left\lbrace +0.62023, -0.66570, +0.68072, -0.68631, +0.68877\right\rbrace,
\end{equation}
as obtained from the Gram-Schmidt procedure.

\section{Outer functions}
\label{app:outerfunction}

The modulus squares of the outer functions for the different form factors are
\begin{align}
    |\phi_{f_t^V}(z)|^2 &=  \frac{(\mLamB -\mLam)^2}{16\pi^2 \chi_V^{J=0}(Q^2)\big|_\text{OPE}} \left( \frac{\sqrt{\lambda(\mLamB^2,\mLam^2,t) }}{t^2 (t-Q^2)^{n+1}}  s_{+}(t) \right)_{t = t(z(\alpha))} \left|\frac{\text{d}z(\alpha)}{\text{d} \alpha} \frac{\text{d}t(z)}{\text{d}z} \right| \,, \label{eq::outerfunc-V-t}\\
    |\phi_{f_0^V}(z)|^2 &=  \frac{(\mLamB +\mLam)^2}{48\pi^2 \chi_V^{J=1}(Q^2)\big|_\text{OPE}} \left( \frac{\sqrt{\lambda(\mLamB^2,\mLam^2,t) }}{t^2 (t-Q^2)^{n+1}}  s_{-}(t) \right)_{t = t(z(\alpha))} \left|\frac{\text{d}z(\alpha)}{\text{d} \alpha} \frac{\text{d}t(z)}{\text{d}z} \right| \,, \label{eq::outerfunc-V-0}\\
    |\phi_{f_\perp^V}(z)|^2 &=  \frac{1}{24\pi^2 \chi_V^{J=1}(Q^2)\big|_\text{OPE}} \left( \frac{\sqrt{\lambda(\mLamB^2,\mLam^2,t) }}{t (t-Q^2)^{n+1}}  s_{-}(t) \right)_{t = t(z(\alpha))} \left|\frac{\text{d}z(\alpha)}{\text{d} \alpha} \frac{\text{d}t(z)}{\text{d}z} \right| \,,\label{eq::outerfunc-V-perp} 
    \end{align}
    \begin{align}
    |\phi_{f_{t}^A}(z)|^2 &= 3 \, |\phi_{f_0^V}(z)|^2   \quad \text{with replacement} \quad \chi_{V}^{J=1}(Q^2)\big|_\text{OPE} \to \chi_{A}^{J=0}(Q^2)\big|_\text{OPE} \,, \\
    |\phi_{f_0^A}(z)|^2 &= \frac{1}{3} \, |\phi_{f_t^V}(z)|^2   \quad \text{with replacement} \quad \chi_{V}^{J=0}(Q^2)\big|_\text{OPE} \to \chi_{A}^{J=1}(Q^2)\big|_\text{OPE} \,, \\
    |\phi_{f_\perp^A}(z)|^2 &=  \frac{1}{24\pi^2 \chi_A^{J=1}(Q^2)\big|_\text{OPE}} \left( \frac{\sqrt{\lambda(\mLamB^2,\mLam^2,t) }}{t (t-Q^2)^{n+1}}  s_{+}(t) \right)_{t = t(z(\alpha))} \left|\frac{\text{d}z(\alpha)}{\text{d} \alpha} \frac{\text{d}s(z)}{\text{d}z} \right| \,,\label{eq::outerfunc-A-perp}
    \end{align}
    \begin{align}
    |\phi_{f_\perp^T}(z)|^2 &=  \frac{ (\mLamB + \mLam)^2}{24\pi^2 \chi_T^{J=1}(Q^2)\big|_\text{OPE}} \left( \frac{\sqrt{\lambda(\mLamB^2,\mLam^2,t) }}{t (t-Q^2)^{n+1}}  s_{-}(t) \right)_{t = t(z(\alpha))} \left|\frac{\text{d}z(\alpha)}{\text{d} \alpha} \frac{\text{d}t(z)}{\text{d}z} \right| \label{eq::outerfunc-T-perp} \\
    |\phi_{f_0^T}(z)|^2 &=  \frac{1 }{48\pi^2 \chi_T^{J=1}(Q^2)\big|_\text{OPE}} \left( \frac{\sqrt{\lambda(\mLamB^2,\mLam^2,t) }}{ (t-Q^2)^{n+1}}  s_{-}(t) \right)_{t = t(z(\alpha))} \left|\frac{\text{d}z(\alpha)}{\text{d} \alpha} \frac{\text{d}t(z)}{\text{d}z} \right| \,,\label{eq::outerfunc-T-0}
    \end{align}
    \begin{align}
     |\phi_{f_\perp^{T5}}(z)|^2 &=  \frac{ (\mLamB - \mLam)^2}{24\pi^2 \chi_{T5}^{J=1}(Q^2)\big|_\text{OPE}} \left( \frac{\sqrt{\lambda(\mLamB^2,\mLam^2,t) }}{t (t-Q^2)^{n+1}}  s_{+}(t) \right)_{t = t(z(\alpha))} \left|\frac{\text{d}z(\alpha)}{\text{d} \alpha} \frac{\text{d}t(z)}{\text{d}z} \right| \,,\label{eq::outerfunc-T5-perp} \\
    |\phi_{f_0^{T5}}(z)|^2 &=  \frac{1}{48\pi^2 \chi_{T5}^{J=1}(Q^2)\big|_\text{OPE}} \left( \frac{\sqrt{\lambda(\mLamB^2,\mLam^2,t) }}{ (t-Q^2)^{n+1}}  s_{+}(t) \right)_{t = t(z(\alpha))} \left|\frac{\text{d}z(\alpha)}{\text{d} \alpha} \frac{\text{d}t(z)}{\text{d}z} \right| \,,\label{eq::outerfunc-T5-0} 
\end{align}
where the value of $\chi_{\Gamma}^{J}(Q^2)\big|_\text{OPE}$ can be found in \reftab{th:db:chiOPE-and-n}.
We can re-express the K\"allen function as $\lambda(\mLamB^2,\mLam^2,t) = s_-(t) s_+(t)$.
Our choice of outer functions $\phi_{f_\lambda^\Gamma}(z)$ must satisfy Eqs.~(\ref{eq::outerfunc-V-t}) - (\ref{eq::outerfunc-T5-0}) and must be analytical within the open unit disk $|z| < 1$. This can be achieved by replacing poles within the unit disk with
\begin{align}
    \left( \frac{1}{t -X}\right)^m \to \left(-\frac{z(t,X)}{t-X} \right)^m\, .
\end{align}
Note that any poles of $1/s_+(t)$ are at $t = (\mLamB + \mLam)^2$, which is mapped by the $z$ transformation to the boundary of the unit disk.
Hence, we do not require any modification to these terms.

Following Refs.~\cite{Boyd:1997qw,Caprini:2019osi}, we compactly express the outer functions of the form factors in a general form:
\begin{align}
    \label{eq:general-outerfunc}
        \phi_{f_\lambda^\Gamma}(z)
            & =
                \frac{\mathcal{N} }{\sqrt{(16+8\cdot c)\cdot d \cdot \pi^2 \chi_\Gamma^{J}\big|_\text{OPE}}} \,
                \phi_1(z)^{e/4} \phi_2(z)^{f/4} \phi_3(z)^{(n+g)/2} \phi_4(z) 
\end{align}
with $\mathcal{N} = (\mLamB + \mLam)^{a} (\mLamB - \mLam)^b$ and 
\begin{align}
    \phi_1(z) &= \left(\frac{s_-(t)}{z(t, (\mLamB-\mLam)^2)} \right) \, , \\
    \phi_2(z) &= s_+(t) \, ,\\ 
    \phi_3(z) &= \left( -\frac{z(t, 0)}{t}\right) \, , \\ 
    \phi_4(z) &= \sqrt{4 (t_+ - t_0)} (1+z)^{1/2}(1-z)^{-3/2} \, . 
\end{align}
The coefficients $a$--$g$ are listed in Table \ref{tab:OuterFunction}.

\begin{table}[t]
    \renewcommand{\arraystretch}{1.25}
    \begin{tabular}{ C{2cm} C{1cm} C{1cm} C{1cm} C{1cm} C{1cm} C{1cm} C{1cm}}
        \toprule
        Outer function  & $a$ &  $b$  & $c$ & $d$ & $e$ & $f$ & $g$\\
        \midrule
        $\phi_{f_t^V}$  & 0 & 1 & 0 & 1 & 1 & 3 & 3 \\ 
        
        $\phi_{f_0^V}$ & 1  & 0 & 1 & 2 & 3 & 1 & 3 \\
        
        $\phi_{f_\perp^V}$  & 0  & 0 & 1 & 1 & 3 & 1 & 2 \\
        
        $\phi_{f_t^A}$  & 1  & 0 & 1 & $\tfrac{2}{3}$ & 3 & 1 & 3 \\
        
        $\phi_{f_0^A}$  & 0 & 1 & 0 & 3 & 1 & 3 & 3 \\ 
        
        $\phi_{f_\perp^A}$  & 0  & 0 & 1 & 1 & 1 & 3 & 2\\
        
        $\phi_{f_0^T}$ & 0  & 0 & 1 & 2 & 3 & 1 & 1\\
        
        $\phi_{f_\perp^T}$  & 1  & 0 & 1 & 1 & 3 & 1 & 2\\
        
        $\phi_{f_0^{T5}}$  & 0  & 0 & 1 & 2 & 1 & 3 & 1\\
        
        $\phi_{f_\perp^{T5}}$  & 0  & 1 & 1 & 1 & 1 & 3 & 2\\
        \bottomrule
    \end{tabular}
    \renewcommand{\arraystretch}{1.0}
    \caption{%
        Summary of the outer functions for each form factor, in terms of the parameters for the general
        decomposition of all outer functions in \refeq{general-outerfunc}.}
    \label{tab:OuterFunction}
\end{table}

\newpage

\section{Convergence}
\label{app:convergence}

We briefly comment on the expected rate of convergence for the series expansion of the
analytic factor of the form-factor parametrization. Consider first the common parametrization in terms of monomials:
\begin{align}
    \text{complete series}
        = \mathcal{P}(q^2) \, \phi_{f_\lambda^\Gamma}(z) f_\lambda^\Gamma(q^2)
        & = \sum_{j=0}^{\infty} b_{j}^{f_\lambda^\Gamma}\, z^j \, .
\end{align}
After truncating this series at order $N$, a change of basis from the monomials to the orthonormal polynomials is possible.
In this way, we have
\begin{align}
  \text{truncated series} = \sum_{j=0}^{N} b_{j}^{f_\lambda^\Gamma} \, z^j = \sum_{i=0}^{N} a_{i}^{f_\lambda^\Gamma}\big|_{N} \, \, p_{i}(z)\,, \label{eq:convergence:a}
\end{align}
with coefficients $a_{i}^{f_\lambda^\Gamma}\big|_{N}$ defined through this equality.\footnote{%
    In practice, the values of these coefficients will of course depend on the statistical approach used to infer
    the form factors from data, which corresponds to known values of the series on the real $z$ axis with $|z| < 1$.
}%
 In the following, we drop the ${f_\lambda^\Gamma}$ superscripts
to enhance legibility.
For any given truncation order $N$, the two sets of coefficients fulfill the linear
relations
\begin{equation}
    a_i\big|_{N} = \sum_{j=0}^{N} X_{ij}\big|_{N} \, b_j\,. \label{eq:Xij}
\end{equation}
The transformation matrix $X$ and the set of coefficients $\lbrace a_i\rbrace|_{N}$ are manifestly dependent on the
choice of truncation order $N$. Note that in general any two sets of coefficients $\lbrace a_i\rbrace|_{N}$
and $\lbrace a_i\rbrace|_{N'}$ are not nested, i.e., the first $\min(N,N')$ elements of the sets are \emph{not}
identical. The coefficients $\lbrace a_{i}\rbrace \big|_{N}$ are bounded by $|a_{i}\big|_{N}|\leq1$, as discussed in the main text.\\

We argue in the following that the coefficients $\lbrace b_{i}\rbrace$ are bounded such that their sum of squares is finite, and the coefficients fall off sufficiently fast to ensure geometric convergence of the truncated series Eq.~\eqref{eq:convergence:a}.
Following Ref.~\cite{Buck:1998kp}, we can use Abel's theorem to show 
that the power series $\sum_j b_j z^j$ converges continuously toward the form factor for $|z| \to 1$ \emph{within the unit disk}.
Making mild assumptions about the asymptotic behaviour of the form factors, in particular smoothness, boundedness, and continuity
on the unit semicircle, Ref.~\cite{Buck:1998kp} suggests that the coefficients $b_j$ are bounded through their $2$-norm
\begin{align}
    \label{eq:app:naive-bound}
    B \equiv \sum_{j=0}^\infty |b_j|^2
\end{align}
in the case of meson-to-meson form factors.
To apply the reasoning of Ref.~\cite{Buck:1998kp} to our case of baryon-to-baryon form factors, we need to focus only on the behaviour for $z \to -1$; all other considerations are
identical.

At $z = -1$ the $\Lambda_b \to \Lambda$ form factors develop a discontinuity due to on-shell intermediate states. Using one of the vector form factors
as an example, $f_\perp^V$ develops a branch cut at $z(t_+) = -1$, where $t_+ = (M_B + M_K)^2$, due to an on-shell
$\bar{B}K$ pair. Other branch cuts appear further along the unit circle.
Watson's theorem provides that the discontinuity of $f_\perp^V(q^2)$ corresponds to the discontinuity of the $B\to K$ form factors
up to a real-valued analytic function in $q^2$, which must be free of kinematic zeros. This holds below the onset of the next branch cut.
As a consequence, the arguments in Ref.~\cite{Buck:1998kp} apply equally to baryon-to-baryon form factors, i.e.,
\begin{equation}
    \sum_{j=0}^\infty b_j < \infty \qquad \text{ and }
    \sum_{j=0}^\infty b_j (-1)^j < \infty
\end{equation}
hold also for the form factors considered here. Moreover, by Parseval's theorem
\begin{equation}
    B = \int_{-\pi}^{+\pi} d\alpha |\phi \mathcal{P} f|^2\bigg|_{z = e^{i\alpha}} = \sum_{j=0}^\infty |b_j|^2 < \infty\,.
\end{equation}
We therefore find that the series on the left-hand side of \eqref{eq:convergence:a} converges rapidly
as $N \to \infty$, and hence so does the series on the right-hand side.\\

This statement can be made more quantitative by considering the truncation error for $|z|<1$, i.e.,
\begin{align}
   \nonumber \text{truncation error}
        & = \Big| (\text{complete series})-(\text{truncated series})\Big| \\
        & = \left|\sum_{j=N+1}^{\infty} b_j \, z^j\right|\, .
\end{align}
Let us write
\begin{equation}
 B^\prime_N \equiv \sum_{j=N+1}^\infty |b_j|^2 \leq B < \infty.
\end{equation}
This implies
\begin{equation}
|b_j|\leq \sqrt{B^\prime_N}\hspace{4ex}\text{for all}\hspace{1ex}j\geq N+1.
\end{equation}
Using the triangle inequality, we find
\begin{align}
   \text{truncation error}
        & \leq \sum_{j=N+1}^{\infty} |b_j \, z^j| \\
        & \leq \sqrt{B^\prime_N} \sum_{j=N+1}^{\infty} |z|^j = \sqrt{B^\prime_N} |z|^{N+1}\sum_{j=0}^{\infty} |z|^j  = \sqrt{B^\prime_N}  \frac{|z|^{N+1}}{1-|z|}\, ,
\end{align}
where $B^\prime_N$ is bounded by the finite constant $B$ and also tends to zero for $N\to\infty$. For our choices of $t_+$ and $t_0$, the $\Lambda_b\to\Lambda$ semileptonic region corresponds to $0\leq z\lesssim 0.23$.
The truncation error in this region thus decreases at least as fast as the geometric sequence $0.23^{N+1}$.\\

\bibliographystyle{jhep} 
\bibliography{main.bib}

\end{document}